\documentclass[prx,amsmath,amssymb, notitlepage, twocolumn,
nofootinbib,
superscriptaddress,
longbibliography
]{revtex4-1}

\usepackage{amsmath}
\usepackage{tabularx,graphicx}
\usepackage{epstopdf}
\usepackage{graphicx}
\usepackage{latexsym}
\usepackage{amssymb}
\usepackage{amsmath}
\usepackage{color, colortbl}
\usepackage{psfrag}
\usepackage{bbm}
\usepackage{bm}
\usepackage{titlesec}
\usepackage{dsfont}
\usepackage{feynmp}
\usepackage{slashed}
\usepackage{multirow}
\newcommand{\mb}[1]{\mathbf{#1}}
\newcommand{\mc}[1]{\mathcal{#1}}

\newcommand{\beq}{\begin{eqnarray}}
\newcommand{\eeq}{\end{eqnarray}}

\newcommand{\la}{\langle}
\newcommand{\ra}{\rangle}

\newcommand{\Tr}{{\rm Tr}}

\newcommand{\bsp}{\begin{split}}
\newcommand{\esp}{\end{split}}

\newcommand{\sgn}{{\rm sgn}}
\newcommand{\eff}{{\rm eff}}
\newcommand{\hc}{{\rm h.c.}}

\newcommand{\ie}{{i.e., }}
\newcommand{\eg}{{e.g., }}
\newcommand{\gcs}{{\rm{CS}}_g}
\newcommand{\mO}{{\mathcal{O}}}
\newcommand{\BCS}{{\rm BCS}}
\newcommand{\DMT}{{\rm DMT}}

\newcommand{\DMMT}{{\rm DM$^2$T }}

\usepackage{color}
\definecolor{darkblue}{rgb}{0.,0.,0.4}
\definecolor{darkred}{rgb}{0.5,0.,0.}
\definecolor{BlueViolet}{RGB}{138,43,226}
\definecolor{SkyBlue}{RGB}{30,144,255}
\definecolor{DarkGreen}{RGB}{0,100,0}
\usepackage[pdftex,colorlinks=true,linkcolor=darkblue,citecolor=blue,urlcolor=darkred]{hyperref}
\usepackage[normalem]{ulem}

\renewcommand{\vec}[1]{\bm{#1}}

\begin{document}
\title{Deconfined metallic quantum criticality: A $U(2)$ gauge-theoretic approach}
\author{Liujun Zou}
\affiliation{Perimeter Institute for Theoretical Physics, Waterloo, Ontario N2L 2Y5, Canada}

\author{Debanjan Chowdhury}
\affiliation{Department of Physics, Cornell University, Ithaca, New York 14853, USA}

\begin{abstract}

We discuss a new class of quantum phase transitions --- Deconfined Mott Transition (DMT)--- that describe a continuous transition between a Fermi liquid metal with a generic electronic Fermi surface and an electrical insulator without Fermi surfaces of emergent neutral excitations. We construct a unified U(2) gauge theory to describe a variety of metallic and insulating phases, which include Fermi liquids, fractionalized Fermi liquids (FL*), conventional insulators and quantum spin liquids, as well as the quantum phase transitions between them. Using the DMT as a basic building block, we propose a distinct quantum phase transition --- Deconfined Metal-Metal Transition (DM$^2$T)--- that describes a continuous transition between two metallic phases, accompanied by a jump in the size of their electronic Fermi surfaces (also dubbed a `Fermi transition'). We study these new classes of deconfined metallic quantum critical points using a renormalization group framework at the leading nontrivial order in a controlled expansion, and comment on the various interesting scenarios that can emerge going beyond this leading order calculation. We also study a U(1)$\times$U(1) gauge theory that shares a number of similarities with the U(2) gauge theory and sheds important light on many phenomena related to DMT, DM$^2$T and quantum spin liquids.

\end{abstract}

\maketitle
\setcounter{tocdepth}{1}
\tableofcontents

\section{Introduction} \label{sec: intro}

Studying the zoo of quantum phases, realized as ground states of many-body systems, and the phase transitions between them is one of the central themes of physics. Quantum phase transitions (QPT) offer an interesting window not just into the universal low-energy physics associated with the criticality, but also on the intricate interplay between the relevant degrees of freedom in the nearby phases. For conventional QPT in insulating phases of matter \cite{Sachdev_book2011}, the critical theory can be described using the classic Landau-Ginzburg-Wilson paradigm developed originally for studying classical (thermal) phase transitions and formulated in terms of the long-wavelength/low-energy fluctuations of a local order-parameter field. A striking example of QPT in insulators that does not fit into this paradigm is offered by the example of `deconfined' quantum criticality \cite{Senthil2003b, Senthil2003a}.

Metallic quantum criticality comes in many {\it avatars} and they are inherently more complicated to describe due to the abundance of gapless electronic excitations near the Fermi surface (FS). Arguably one of the most common forms of metallic criticality is associated with the onset of some form of broken-symmetry in a metal, which is typically described in terms of the critical fluctuations of a bosonic order-parameter coupled to a Fermi surface. There are numerous strongly-correlated materials where such forms of criticality might be relevant for the experimental phenomenology \cite{rmpqcp,Stewart,Matsuda14}. The standard Hertz-Millis-Moriya (HMM) framework \cite{rmpqcp} for dealing with this problem runs into serious difficulties in (2+1)-dimensions \cite{Lee2009}; much theoretical progress has nevertheless been made in recent years \cite{Leereview}.

One of our main focus in this paper will be on continous metal-insulator transitions. In this context, there are simpler and relatively well understood mechanisms. For example, in clean systems, an example of such a QPT is through a Lifshitz transition into a band insulator, where the electronic Fermi surface can be shrunk to a point (e.g. by changing the chemical potential). In disordered systems, a localization transition into an Anderson insulator can localize the electronic states near the Fermi level in the presence of strong disorder.

A conceptually distinct and somewhat novel form of metallic criticality is associated with the disappearance of an entire Fermi surface as a result of either a Kondo breakdown transition in heavy Fermi liquids \cite{Coleman2000,Colemanetal,Si1,Si2}, or, a Mott transition to a quantum spin liquid insulator \cite{Senthil2003,Senthil2008}. In both of these examples, a description of the transition requires the introduction of fractionalized degrees of freedom (d.o.f.) coupled to an emergent dynamical gauge field, and the electronic Fermi surface evolves into a {\it ghost} Fermi surface of the electrically neutral fractionalized d.o.f. across the transition. The resulting critical point has a `critical' Fermi surface---a sharp Fermi surface of electrons without any long-lived low-energy electron-like quasiparticles. See Refs. \cite{Senthil2003,Senthil2008,TS08,CFL_FL,DChiggs} for some examples of such transitions. A key question that remains unanswered is as follows: {\it Can we describe continuous quantum phase transitions from a metal to an insulator (or to another distinct metal) without any remnant Fermi surfaces of fractionalized d.o.f. on the insulating side (or the other metallic side)?}

In this paper, we introduce a new gauge-theoretic formulation to study the exotic quantum phase transitions out of a metal introduced above. As already emphasized, our main focus will be on a distinct class of continuous QPT between a Fermi liquid (FL) metal and a Mott insulator (MI), where the size of the electronic Fermi surface is finite on one side of the transition within the FL phase \footnote{This is different from a Lifshitz-type transition, where the size of the Fermi surface on the metallic side is infinitesimal immediately across the transition.}, and the MI has \underline{no} FS of any fractionalized d.o.f.  We dub such metal-insulator transitions as {\it Deconfined Mott Transition} (DMT). Notice that the insulating phase across the DMT can either be short-range entangled (\eg a conventional magnet) or long-range entangled (\eg a quantum spin liquid (QSL)), and the only requirement of the insulator is the absence of any ghost Fermi surfaces. We note, in passing, that numerical evidence for a DMT between a FL and a chiral spin liquid has been reported recently \cite{Szasz2018}. \footnote{In the context of semi-metals (rather than metals), transitions of a somewhat similar flavor between a Dirac semi-metal and a {\em featureless} insulator have been discussed \cite{You2017, You2017a}.}

One of our motivations to study DMT is clearly their conceptual novelty. From a more practical point of view, one of our other motivations comes directly from experiments on numerous strongly correlated materials, where the same framework for a DMT may offer us a useful route to understanding continuous QPT between different metals. In numerous rare-earth element based compounds that display heavy-fermion physics, the Kondo-breakdown transition out of the heavy FL into a metal with broken symmetries display phenomenology at odds with the expectations of HMM theory (see discussion in Ref.~\cite{Senthil2003} for some examples). Similarly, in the hole-doped cuprates, there is growing evidence for a transition between a FL metal at large doping to an unconventional `pseudogap' metal with a markedly different density at small values of the doping (without any additional exotic FS); see Ref.~\cite{Keimer15} and references therein. Inspired by these examples, we are interested in studying a direct continuous QPT between two metals that have electronic FS with finite but different sizes on either side of the transition. We dub such QPT as {\it Fermi Transition} (FT), or, {\it Deconfined Metal-Metal Transition} (DM$^2$T).  We note that within our organizing principle, the two metallic phases separated by a \DMMT can either be ordinary FL metals, or be of a more exotic nature, such as a fractionalized Fermi liquid (FL*) \cite{Senthil2002}. 

We remark that in discussing DMT and DM$^2$T, we require that immediately across the transition, the total size of the metallic FS is finite, which is allowed to be a multiple of the size of the Brillouin zone (BZ). Later we will see an explicit example of a metal that has two FS whose total size is the same as the BZ size. The DMT or DM$^2$T out of such a metallic phase possesses all of the essential characteristics of related transitions in metals where the FS size is not a multiple of the BZ size.

In the remainder of our discussion, we exclude electronic transitions purely driven by the onset of various forms of translation-symmetry-breaking order, such as spin or charge density waves \cite{SS95}. The latter can be described within the conventional HMM framework. Infinitesimally close to the critical point on the ordered side of such continuous transitions (\ie when the order-parameter is infinitesimally small in magnitude), the Fermi surfaces are nearly identical to the one on the disordered side of the transition, except for in the narrow vicinity of the `hot-spots'. We are interested in DM$^2$T, where the electronic Fermi surface undergoes an abrupt change across the critical point, which necessarily goes beyond the HMM framework.

\begin{figure}
	\centering
	\includegraphics[width=\linewidth]{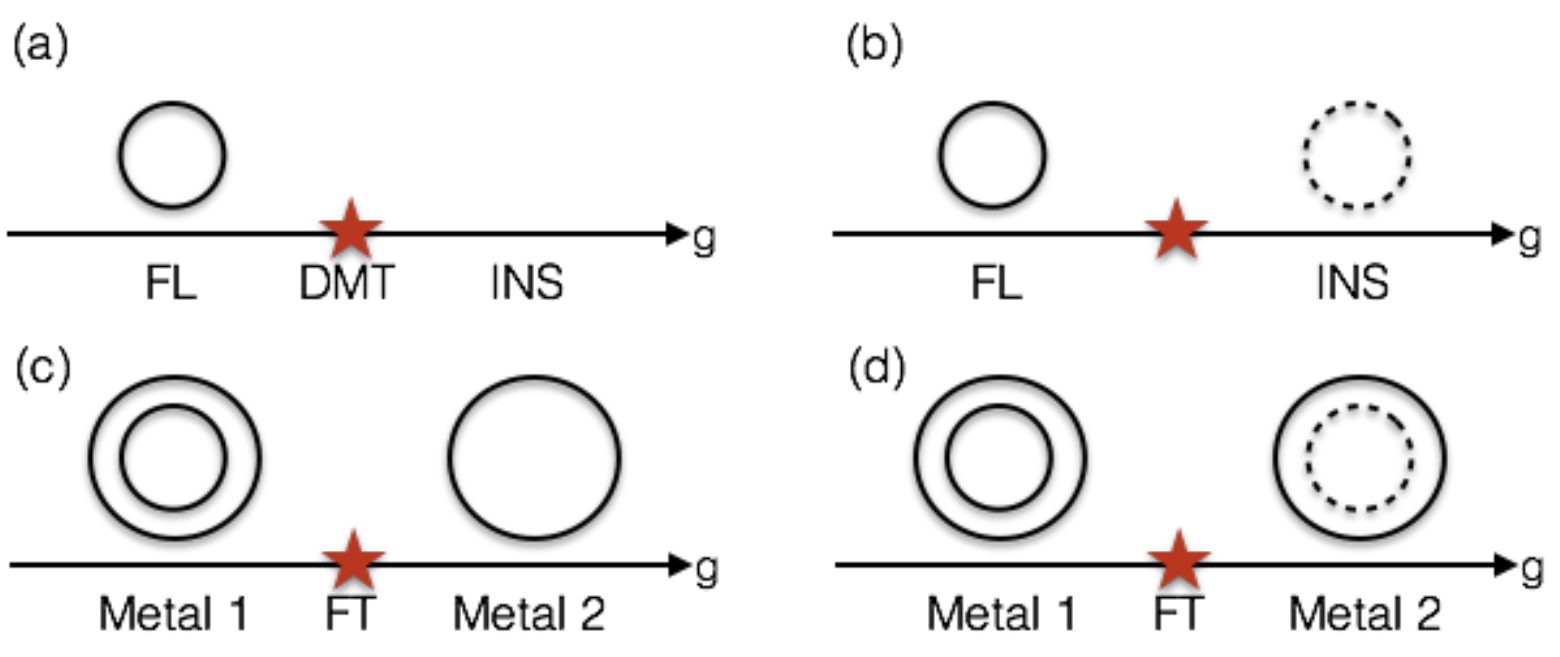}
	\caption{Examples of deconfined metallic quantum criticality: (a) deconfined Mott transition (DMT) between a Fermi liquid (FL) and an insulator without a ghost Fermi surface. The solid circle represents an electronic Fermi surface. (b) Mott transition between a FL and an insulator with a ghost Fermi surface, as described in Refs. \cite{Senthil2008}. The dashed circle represents a neutral ghost Fermi surface. (c) A Fermi transition (FT)  between two distinct FL metals with FS of different size (also dubbed as deconfined metal-metal transition (\DMMT)). The FL on the left has FS of `hot' electrons (which undergo a DMT as in (a)) and `cold' electrons (which remain spectators). (d) a FT between a FL and a fractionalized Fermi liquid (FL*), driven by the hot electrons undergoing a transition into a neutral ghost Fermi surface, as in (b) (see Ref. \cite{Senthil2003}).}
	\label{fig: transitions}
\end{figure}

It might already be apparent to the reader that the framework for a DMT can be readily generalized to describe a \DMMT. For example, consider a two-orbital electronic model with a globally conserved density, where each orbital has a FS. We denote these FS as `hot' and `cold', respectively, for reasons that will be clear momentarily. From Luttinger's theorem, the total size of the FS is the sum of the sizes of the individual FS from the two orbitals. Suppose now that as a result of strong orbital-selective interactions, the hot electrons undergo a DMT, while the cold electrons remain in a FL phase with a fixed FS size. If the coupling between the hot and cold electrons is irrelevant in a renormalization-group (RG) sense, the cold electrons decouple and act as spectators to the more interesting DMT of the hot electrons. Therefore, such a transition can be viewed as a \DMMT of the entire system. It is worth emphasizing that this is only one possible route to a \DMMT, and it does not imply that all \DMMT can necessarily be viewed as a DMT on top of a dynamically decoupled FL. A schematic representation of the different metallic QPT of interest to us in this paper appears in Fig. \ref{fig: transitions}.

\begin{figure}
	\centering
	\includegraphics[width=0.9\linewidth]{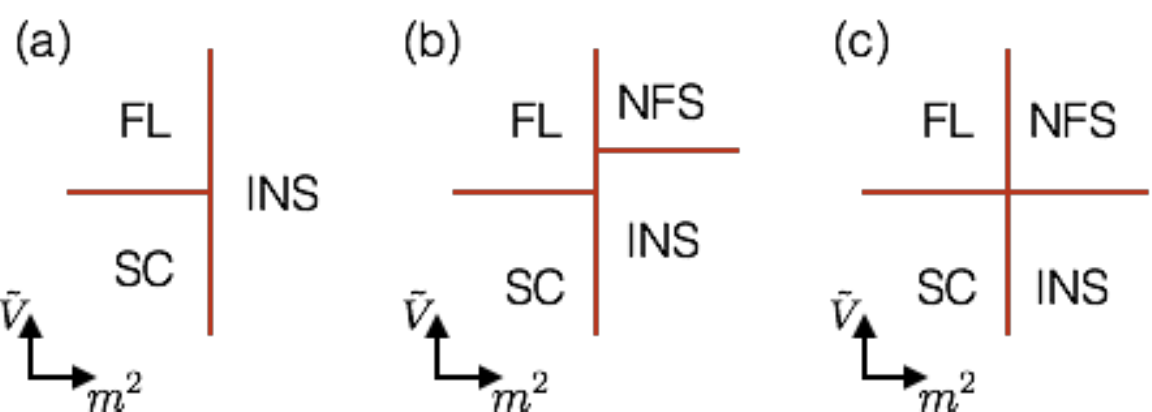}
	\caption{Some possible landscapes of the allowed phases. The parameter $m^2$ indicates the gap of the bosonic parton (see Eq. \eqref{eq: boson-gauge action}), and $\widetilde V$ represents the short-range interaction among the fermionic partons (see Sec. \ref{subsubsec: pairing instability}). Here FL, SC, INS and NFS stand for a Fermi liquid metal, a superconductor, an insulator without a `ghost' Fermi surface of fractionalized d.o.f., and a quantum spin liquid with an emergent neutral Fermi surface, respectively. The red lines are schematic phase boundaries. A direct DMT without fine-tuning can occur only in case (a) and (b).}
	\label{fig: landscape}
\end{figure}

The focus of this paper is to provide a possible mechanism for a DMT, and a related mechanism for a DM$^2$T. Our key strategy will be to formulate the problem in terms of an (emergent) $U(2)$ gauge theory. More precisely, we will view the electron as a bound state of a boson and a fermion, where the boson carries all of the quantum numbers of the electron and the fermion carries only the fermionic statistics. Furthermore, both the boson and the fermion are coupled to a dynamical $U(2)$ gauge field. We will show that by tuning a few parameters of the resulting low-energy theory, this powerful unified framework is capable of describing multiple quantum phases, including FL metals, orthogonal metals, ordinary insulators, different topologically ordered states and FL* metals. The same framework will offer new insights into the putative critical theories for a DMT and a \DMMT, which are described by a theory of coupled critical bosons, fermions with FS, a dynamical $U(2)$ gauge field, and possibly a FS of cold electrons. A landscape of the possible phases and the QPT between them in terms of some of the parameters in our theory appears in Fig. \ref{fig: landscape}. Exploring the nature of all of the QPT that appear on Fig. \ref{fig: landscape} is beyond the scope of the present manuscript and will appear in future work.

We apply a renormalization group (RG) analysis to critically examine whether these putative critical theories describe continuous DMT and DM$^2$T. We find that the $U(2)$ gauge field is effectively `quasi-Abelianized', \ie many of its non-Abelian features do not actually play an important role in the universal low-energy limit that is of physical interest to us. Our calculations provide considerable insight and some preliminary evidence that the sought-after QPT can be described by these critical field theories. However, we stress that corrections beyond those considered here can potentially alter the nature of these transitions. A more elaborate investigation in the future is required to fully understand the intricate low-energy dynamics of these theories. 

Along the way, we also study a $U(1)\times U(1)$ gauge theory that shares many common aspects as the $U(2)$ gauge theory. As we discuss, our results on this $U(1)\times U(1)$ gauge theory have important implications on various interesting phenomena related to DMT, DM$^2$T and QSLs.

The rest of this paper is organized as follows: In Sec. \ref{sec: U(2) gauge theory} we provide a concrete parton construction that naturally leads to a $U(2)$ gauge theory. We discuss the routes to obtaining different phases, including a metallic FL, various insulating phases (including short-range entangled insulators and long-range entangled insulators) and metallic FL* phases starting with the above $U(2)$ gauge theory. In Sec. \ref{sec: transitions}, we discuss possible mechanisms for a DMT and a \DMMT based on the $U(2)$ gauge theory. In Sec. \ref{sec: RG}, we apply a renormalization group analysis to the $U(2)$ gauge theory of a DMT, as well as a similar $U(1)\times U(1)$ gauge theory. While we will provide preliminary and compelling evidence that the theory can describe a continuous transition of this type, there are technical challenges that prevent us from addressing some aspects of this transition. Finally, in Sec. \ref{sec: discussion}, we conclude with a discussion of the possible scenarios for the global landscape of the underlying $U(2)$ gauge theory.

\section{$U(2)$ gauge theory} \label{sec: U(2) gauge theory}

In this section we present a formalism based on a $U(2)$ gauge theory and apply it to describe various quantum phases of interest to us. We will see that within our framework these phases can be obtained by tuning a few parameters. {\footnote{We note in passing that $U(2)$ gauge theories have been studied recently in a different context in other interesting settings \cite{Zou2018}, but our current framework is different from the previous $U(2)$ gauge theory.}}

\subsection{Parton construction} \label{subsec: parton construction}

We start by presenting a parton decomposition of the electron:
\beq \label{eq: parton}
\left(
\begin{array}{c}
c_1\\
c_2
\end{array}
\right)
=
\left(
\begin{array}{cc}
B_{11} & B_{12} \\
B_{21} & B_{22}
\end{array}
\right)
\cdot
\left(
\begin{array}{c}
     f_1  \\
     f_2
\end{array}
\right)
\eeq
or, more schematically,
\beq \label{eq: parton schematic}
c=Bf,
\eeq
where $c_1~(c_2)$ is the annihilation operator of an electron with spin-up (-down). The $B$'s are bosonic operators and the $f$'s are fermionic. This parton construction has a $U(2)$ gauge redundancy, under which $f\rightarrow Uf$ and $B\rightarrow BU^\dag$, with $U$ a $U(2)$ matrix. This means that at low energies, the system is described by a theory where the partons, $B$ and $f$, are coupled to a dynamical $U(2)$ gauge field, denoted by $\mb a$. We will refer to the index of $f$ and the second index of $B$ as the color index, while the first index of $B$ is the physical spin index.

We leave details of this parton construction to Appendix \ref{app: U(2) parton}, where we discuss the generators of the $U(2)$ gauge transformations, the gauge constraints of a physical state and the procedure for constructing physical electronic states using $B$ and $f$. Below we only summarize the actions of the relevant physical symmetries that we will use in this paper.

Under the $U(1)$ charge conservation symmetry,
\beq
B(\vec r)\rightarrow e^{i\theta}B(\vec r),
\quad
f(\vec r)\rightarrow f(\vec r).
\eeq
Under the $SU(2)$ spin rotation symmetry,
\beq
B(\vec r)\rightarrow VB(\vec r),
\quad
f(\vec r)\rightarrow f(\vec r),
\eeq
where $V$ is an $SU(2)$ matrix. Under time reversal $\mc{T}$,
\beq
B(\vec r)\rightarrow \epsilon B(\vec r),
\quad
f(\vec r)\rightarrow f(\vec r),
\quad
i\rightarrow -i,
\eeq
where $\epsilon$ is the rank-2 anti-symmetric matrix with $\epsilon_{12}=-\epsilon_{21}=1$. Under lattice symmetries,
\beq
B(\vec r)\rightarrow B(\vec r'),
\quad
f(\vec r)\rightarrow f(\vec r')
\eeq
where $\vec r'$ is the transformed position of $\vec r$ under the relevant lattice symmetry. {\footnote{Notice the above is only a particular choice of how symmetries act on $B$'s and $f$'s, and in general there are different projective representations of these symmetries \cite{Wen2004Book}, which will not be discussed in this paper.}}

The parton construction given by Eq. \eqref{eq: parton} can be generalized to a many-body system in a straightforward fashion. In terms of these partons and the emergent $U(2)$ gauge field, the low-energy dynamics of the system can be described by an action of the following form:
\beq \label{eq: theory schematic}
S_{\DMT}=S_{[B, \mb{a}]}+S_{[B,f]}+S_{[f, \mb{a}]}+S_{[\mb{a}]},
\eeq
where $S_{[B, \mb{a}]}$ ($S_{[f, \mb{a}]}$) represent the actions of $B$ ($f$) that are both minimally coupled to $\mb{a}$, $S_{[B, f]}$ represents the coupling between $B$ and $f$, and $S_{[\mb{a}]}$ represents the action of the $U(2)$ gauge field, $\mb{a}$. In the above expressions, the couplings to external gauge fields corresponding to the global symmetries are implicitly displayed, according to the symmetry actions specified above. The precise forms of these actions are unimportant at this stage, and they will be specified later.

Below we apply this $U(2)$ gauge theory to describe different quantum phases.

\subsection{Metallic phases} \label{subsec: metals}

The above parton construction can be applied to describe a Fermi liquid in any dimension ($>1$) and on any lattice geometry. For concreteness, we consider a two dimensional triangular lattice with two electronic orbitals per site at half-filling (\ie two electrons per site on average). Throughout this paper, we will assume that the mean-field part of the fermion action, $S_{[f, \mb{a}]}$, corresponds to a Hamiltonian that gives rise to a generic FS (\ie a FS without any special nesting property). A particular choice of the mean field Hamiltonian for $f$ is
\beq
\begin{split}
H_f=
&-t_1\sum_{\la\vec r\vec r'\ra}\sum_{\alpha, i=1,2}f_{\alpha i}^\dag(\vec r) f_{\alpha i}(\vec r')\\
&-t_2\sum_{\vec r}\sum_{\alpha=1,2}f_{\alpha 1}^\dag(\vec r)f_{\alpha 2}(\vec r)
+\hc,
\end{split}
\eeq
where $\alpha=1,2$ represent the color indices, and $i=1,2$ represent the two orbitals. The first term above represents the nearest-neighbor hopping within the same orbital, and the second term represents the hybridization between the two orbitals at the same site. When $|t_2|\gtrsim|t_1|$, there is no FS. However, when $|t_1|\gtrsim |t_2|$, at half filling, there will be two FS sheets with different sizes that add up to the size of the BZ. We will focus on the latter case here. It is possible to have a metal-insulator transition by tuning $t_1/t_2$, without changing the total size of the Fermi surface up to the BZ size. However, later we will discuss a more exotic possibility of metal-insulator transition, \ie a DMT where on the metallic side the sizes of the two FS are finite in the immediately vicinity of the transition.

The bosons, $B$, can be condensed or gapped out by tuning appropriate parameters in $S_{[B, \mb{a}]}$. If $B$ is condensed, the dynamical $U(2)$ gauge field gets Higgsed generically and does not play an important role at low energies.
According to the original relation in Eq. \eqref{eq: parton}, the FS of $f$ are then identified with the FS of the original electron with a renormalized quasiparticle residue, $Z\sim\la |B|^2\ra$. The resulting state is simply a Fermi liquid metal. Furthermore, by choosing appropriate interaction terms in $S_{[B, \mb{a}]}$, $B$ can be condensed in a channel where all physical symmetries are preserved \cite{Seiberg2016, Zou2018}. For example, the condensate of $B$ can have the form $\la B_{11}\ra=\la B_{12}\ra=\la B_{21}\ra=-\la B_{22}\ra=B_0$, where $B_0\neq 0$ is independent of the site or orbital. It is straightforward to check that this condensate is invariant under all physical symmetries discussed above (up to a $U(2)$ gauge transformation).

Finally we note that it is straightforward to realize other metallic phases in this formalism. For example, by choosing the parameters in $S_{[B, \mb{a}]}$ such that $B$ is condensed in a channel with $\la B_{11}\ra=2\la B_{22}\ra=B_0\neq0$ and $\la B_{12}\ra=\la B_{21}\ra=0$ realizes a ferromagnetic FL metal if $B_0$ is independent of the site or orbital index. On the other hand, if $B_0$ oscillates from site to site, it realizes a spin-density wave metal. As another example, when $B$ is gapped but a bound state of $B$ is condensed, \eg by taking $\la B\ra=0$, $\la B_{11}^2\ra=\la B_{22}^2\ra\neq 0$ and $\la B_{12}^2\ra=\la B_{21}^2\ra=0$, the $U(2)$ gauge structure can be broken down to $Z_2$, which realizes an orthogonal metal \cite{Nandkishore2012}.

Therefore, the above $U(2)$ gauge theory can describe a wide variety of metallic states, depending on whether $B$ or its bound states are condensed in an appropriate fashion.

\subsection{Insulating phases} \label{subsec: insulators}

Having constructed various metallic phases from the above $U(2)$ gauge theory by condensing $B$, we are now interested in analyzing the phases that one obtains when $B$ is gapped while the mean field Hamiltonian for $f$ describes a generic FS. The nature of the resulting state is determined by the state $B$ is in, as well as the interaction between both partons and the dynamical $U(2)$ gauge field, $\mb{a}$. In order to approach this problem in a systematic fashion, below we first ignore the coupling of $B$ to $f$ and $\mb{a}$, and discuss the possible gapped states that $B$ can be in. We then combine each of these states with the $f$ and $\mb{a}$ to address the possible insulators that can be accessed by the $U(2)$ gauge theory.

Generically, there can be the following four types of gapped states of $B$. Notice that in all of the gapped states that we will discuss here, the global symmetries of the system may or may not be broken spontaneously, depending on the microscopic details (which we do not specify). For our particular setup of the lattice problem, none of the symmetries has to be broken spontaneously. However, depending on the particular setup of the lattice problem, there can be a Lieb-Schultz-Mattis (LSM) type theorem that forbids a featureless state of $B$ \cite{Lieb1961, Affleck1988, Oshikawa1999, Hastings2003}. The general discussion below applies to all lattice setups, and we will explicitly point out the features that are specific to our particular setup.

\begin{enumerate}
    
    \item Type-I: A short-range entangled (SRE) state with a trivial response to the dynamical $U(2)$ gauge field, $\mb{a}$. A particular example of such a state is where the $B$'s within the two orbitals at each site form a dimer:
    \beq
    \prod_{\vec r}\epsilon_{\alpha\alpha'}\epsilon_{\beta\beta'}(\sigma_x)_{ij}B^\dag_{\alpha\beta i}(\vec r)B^\dag_{\alpha'\beta' j}(\vec r)|0\ra,
    \eeq
    where $|0\ra$ represents a state with no bosons, $i, j=1,2$ represent the two orbitals at a site, and $\vec r$ labels the position of the site. This is a product state that preserves all global symmetries as well as the $U(2)$ gauge symmetry. Notice the existence of such a featureless state is due to our particular choice of lattice. If we were to consider instead another lattice system with a LSM-type constraint, the SRE state needs to spontaneously break some symmetry.
    
    \item Type-II: A SRE state with a nontrivial response to the dynamical $U(2)$ gauge field, $\mb{a}$. That is, $B$ forms a nontrivial symmetry-protected topological (SPT) state under the $U(2)$ gauge symmetry. It is known that in two dimensions if $B$ is in such a nontrivial SPT, integrating $B$ out generates a Chern-Simons action for $\mb{a}$ \cite{Liu2012, Senthil2012, Seiberg2016, Zou2018, Ning2019}, which can potentially modify the low-energy dynamics of the entire system.
    
    \item Type-III: A long-range entangled (LRE) state without a nontrivial Chern-Simons response to the dynamical $U(2)$ gauge field, $\mb{a}$. An example of such a state is a $Z_2$ topological order \cite{Wen2004Book}. It will be useful later on to notice that since $B$ is already in the fundamental representation of the $SU(2)$ gauge symmetry, the $SU(2)$ gauge charge of the topological order can be completely screened by $B$, \ie the sector of the topological order can always be made into a singlet sector of the $SU(2)$ gauge symmetry.
    
    \item Type-IV: A LRE state with a nontrivial Chern-Simons response to the dynamical $U(2)$ gauge field, $\mb{a}$. Such a state can be obtained by forming a nontrivial SPT of $B$ on top of a type-III state of $B$. 
    
\end{enumerate}

Next we discuss the dynamics of the $f$ and $\mb{a}$ sector. According to the above discussion, in this case, at low energies the system is described by FS of $f$ that are coupled to the dynamical $U(2)$ gauge field, $\mb{a}$. The $U(2)$ gauge field can be decomposed as $\mb{a}=a+\tilde a\mb{1}$, where $a$ is a 2-by-2 $SU(2)$ gauge field, and $\tilde a$ is a $U(1)$ gauge field. The meaning that $\mb{a}$ is a $U(2)$ gauge field is that an excitation with an odd (even) charge under $\tilde a$ must carry a half-odd-integer (integer) spin under $a$, and vice versa. Generically, there is no relation between the $SU(2)$ gauge coupling and the $U(1)$ gauge coupling---this will be an important consideration for us later.

The leading instability of the $f$-FS will be to pairing. In our setting, as a result of the difference in the size of the two FS, pairing is likely to occur within the same FS. The coupling between $f$ and $\mb{a}$ is expected to have significant influence on the pairing instability. However, independent of the details of this coupling (which we will discuss in great detail later in Sec. \ref{sec: RG}), we can always explicitly add a four-fermion attractive interaction to the theory in Eq. \eqref{eq: theory schematic} so that there is an $s$-wave pairing in the singlet channel of the $SU(2)$ gauge symmetry. The $f$-FS is then fully gapped out and the $U(2)$ gauge structure will be broken down to an $SU(2)$ gauge structure. The nature of the final state then depends on the dynamics of the remaining $SU(2)$ gauge field. The advantage of studying the fate of the theory in the absence of gapless FS of the $f$ is that the dynamics of the remaining $SU(2)$ gauge field only depends on the nature of the insulating state $B$ is in (\ie Type-I,..,IV). We discuss these in detail below. We shall return to the full problem in the presence of the $f$-FS in Sec. \ref{sec: transitions}, where we will also investigate the tendency to pairing of the $f$-fermions.

\begin{enumerate}
    
    \item Type-I: In this case, it is generally believed that the $SU(2)$ gauge field confines \cite{Polyakov1987}. It turns out that the system is in an ordinary SRE insulator (see Appendix \ref{app: type-2}). 
    
    \item Type-II: In this case, the $SU(2)$ gauge field will not confine due to its Chern-Simons action, and the resulting state is generically a topologically ordered state. For example, if the SPT of $B$ is a $U(2)$ symmetric bosonic integer quantum Hall state discussed in Ref. \cite{Senthil2012}, using the method in Ref. \cite{Zou2018} (see Appendix \ref{app: type-2}), we can determine that the resulting topological order is equivalent to a chiral spin liquid state \cite{Wen2004Book}, also known as the Laughlin-$1/2$ state. Notice in the usual terminology, a chiral spin liquid refers to a quantum spin liquid emergent from a {\it bosonic} system, but in our case gapped local electrons are also present in the system.  
    
    \item Type-III: In this case, the $SU(2)$ gauge field will confine. Interestingly, according to the discussion above, because the topological order of $B$ can be made completely in the singlet sector of the $SU(2)$, this topological order is intact when the $SU(2)$ gauge field confines. As a result, the resulting state of the entire coupled system of $B$, $f$ and $\mb{a}$ is again generically a topologically ordered state, whose precise nature is determined by both the topological order of $B$ and the symmetry fractionalization pattern of the $U(2)$ gauge symmetry on this topological order. In Appendix \ref{app: type-2} we give an example where the system is in a $Z_2$ topologically ordered state (supplemented with local electrons).

    \item Type-IV: In this case, the $SU(2)$ gauge field does not confine due to its Chern-Simons action, and the resulting state is also generically a topologically ordered state. In Appendix \ref{app: type-2}, we give a relatively simple example of such a state.
    
\end{enumerate}

In summary, our $U(2)$ gauge theory can describe insulators that are either SRE or LRE, depending on the parameters of Eq. \eqref{eq: theory schematic}. Notice that all of the above insulating phases have no FS of any emergent excitations (as long as the attractive four-fermion interaction is reasonably strong).

In passing, we note that one can also construct gapless phases of $B$ (\eg a Dirac spin liquid \cite{Wen2004Book}) that respect the $U(2)$ gauge symmetry. However, we do not explore such scenarios any further in this paper.

\subsection{Fractionalized Fermi liquids} \label{subsec: FL*}

Our formalism can also describe fractionalized Fermi liquids (FL*) \cite{Senthil2002}, upon including additional d.o.f. in the problem. Consider including  additional electrons that form a FS. Let us denote the creation and annihilation operators for these electrons by $d^\dag,~d$. Then the modified theory is given by 
\beq \label{eq: FT theory schematic}
S_{{\rm DM^2T}}=S_{\DMT}+S_{[d]}+S_{[d,c]},
\eeq
where $S_{\DMT}$ is as originally defined in Eq. \eqref{eq: theory schematic}, $S_{[d]}$ is the action of the $d$-electrons. The coupling between the $d$- and $c$-electrons, $S_{[d, c]}=\int d\tau d^2x\mc{L}_{[d, c]}$, takes the familiar form
\beq \label{eq: cold-hot coupling}
\mc{L}_{[d, c]}=\lambda d^\dag Bf+\hc+\cdots
\eeq
with $\lambda$ the hybridization strength between the $d$- and $c$-electrons. 

If the $c$-electrons form a FL (\ie if $B$ is condensed), the entire system is just a FL made of both the $d$- and $c$-electrons (with a single globally conserved density). On the other hand, when $B$ is gapped, the resulting metallic phases can be potentially interesting. If $B$ is gapped and the $c$-electrons form an insulator of Type-I discussed above, the system can be viewed as a FL of the $d$-electrons on top of an ordinary insulating state of the $c$-electrons. However, if $B$ is gapped and the $c$-electrons form an insulator of Type-II, -III or -IV described above, the system can be viewed as a FL of the $d$-electrons coexisting with an essentially decoupled (in an RG sense) topologically ordered state of the $c$-electrons. In other words, this provides us with a route towards realizing FL*; by changing the nature of the gapped state of $B$, different FL* ground states can be obtained.

To conclude this section, we remark that our single unified framework is able to describe a myriad of interesting quantum phases. They include various metallic phases with or without broken symmetries/topological order as well as fully gapped insulators. The latter are obtained when both $B$ and $f$ are gapped (at least when the attractive interaction between the $f$'s is reasonably strong). Moreover, letting at least one of them be in a nontrivial SPT under the $U(2)$ gauge symmetry allows us to describe many types of topologically ordered phases \cite{Zou2018}. The full technical advantage of our setup will become apparent in the following discussion.

\section{Deconfined Mott and Metal-Metal Transitions} \label{sec: transitions}

Based on the above descriptions of various phases in terms of our $U(2)$ gauge theory, in this section we present possible mechanisms for a deconfined Mott transition and a deconfined metal-metal transition. In this section, we will only restrict ourselves to making rather general remarks on these mechanisms. The quantitative analysis using a formal machinery of renormalization group for these critical theories appears in Sec. \ref{sec: RG}.

\subsection{Deconfined Mott Transition} \label{subsec: DMT mechanism}

In this section, we present a possible mechanism for a DMT between a FL metal and an insulator without a FS of any fractionalized d.o.f., starting from our $U(2)$ gauge theory. For the sake of concreteness, the insulator we focus on here will be SRE (\ie Type-I in Sec. \ref{subsec: insulators}); similar mechanisms can be applied to other types of insulators.

We begin by reminding the readers that in our previous discussion of the possible insulators in Sec. \ref{subsec: insulators}, we assumed a relatively strong short-ranged attractive interaction between the $f$-fermions in the singlet channel of the $SU(2)$ gauge symmetry to induce pairing. However, in this regime across the condensation transition for the boson, a relatively strong short-ranged attraction may be present for the electrons as well, thereby inducing pairing of the FL. Thus, we need to relax the requirement of strong attraction in order to describe a DMT transition from a SRE insulator to a FL metal (rather than to a superconductor). We are thus immediately forced to revisit the question of whether the SRE insulator can be obtained in the regime without strong attraction in the first place (\ie whether the $f$ FS is still unstable to pairing, as a result of the coupling to $\mb{a}=a+\tilde a\mb{1}$, when the $B$'s are gapped)?

It is well known that a $U(1)$ gauge field, $\tilde a$, coupled minimally to a neutral FS suppresses pairing \cite{Metlitski2014}, while the $SU(2)$ gauge field mediates a long-range attractive interaction among the fermions in the $SU(2)$ singlet channel, thereby enhancing the tendency towards pairing. Therefore, in the regime without a strong short-ranged attraction between the $f$-fermions, the fate of the FS to pairing depends on the competition between the $U(1)$ and $SU(2)$ gauge fields. If the $U(1)$ gauge field wins over the $SU(2)$ gauge field and the FS remain stable, we obtain a new type of quantum spin liquid that has stable neutral FS of fractionalized d.o.f. coupled to a dynamical $U(2)$ gauge field. The transition from a FL to such an exotic state is clearly interesting, but it is not an example of a DMT, as the above insulator has a neutral FS. On the other hand, if the $SU(2)$ gauge field wins over the $U(1)$ gauge field and the FS become unstable to pairing, then this approach can potentially describe a DMT from a FL. In Sec. \ref{sec: RG} we will provide suggestive evidence that the $SU(2)$ gauge field wins over the $U(1)$ gauge field in this competition, although our results at this stage cannot account for all of the intricate complexities and much more remains to be understood. In the remainder of this Sec. \ref{sec: transitions}, we assume that the $SU(2)$ gauge field indeed wins over the $U(1)$ gauge field and continue to discuss the associated critical theory. In Sec. \ref{sec: discussion} we will discuss other possible scenarios where the $U(1)$ gauge field wins over the $SU(2)$ gauge field in this competition. We show a few plausible phase diagrams in Fig. \ref{fig: landscape}, where a direct DMT without fine-tuning can only occur in cases (a) and (b).

Based on the above discussion, we now flesh out the details of the boson ($B$) induced order-disorder transition, which potentially describes a DMT between a FL metal and a SRE insulator. The critical theory is described by the action $S_{\DMT}$ in Eq. \eqref{eq: theory schematic}, where $B$ is tuned to be critical. For the transition into a Type-I insulator (with a fixed electron density), $S_{[B, \mb{a}]}$ can be taken to be $S_{[B, \mb{a}]}=\int d\tau d^2x\mc{L}_{[B, \mb{a}]}$ with a relativistic Lagrangian,
\beq \label{eq: boson-gauge action}
\mc{L}_{[B, \mb{a}]}=\Tr\left(D^\mu_{\mb{a}}B^\dag D^\mu_{\mb{a}}B\right)+m^2\Tr(B^\dag B)+\cdots,
\eeq
where $D^\mu_{\mb{a}}=\partial^\mu-i\mb{a}^\mu$ is the covariant derivative with respect to $\mb{a}$ (we have set the boson velocity to unity), $m^2$ is the tuning parameter of the transition, and ``$\cdots$" represents all other terms that are allowed by global symmetries and gauge invariance.

The coupling between $B$ and $f$, $S_{[B, f]}=\int d\tau d^2x\mc{L}_{[B, f]}$, needs to be gauge invariant, so direct coupling between $B$ and bilinears of $f$ is forbidden, and their leading coupling is of the form
\beq
\mc{L}_{[B, f]}=\lambda_1f^\dag B^\dag Bf+\lambda_2 f^\dag f\cdot\Tr(B^\dag B)+\cdots,
\eeq
where $\lambda_1$ and $\lambda_2$ are coupling constants.

The coupling between $f$ and $\mb{a}$ has the form
\beq
S_{[f, \mb{a}]}=\int_{\omega,\vec k}f^\dagger\left(-i\omega-\mu_f+i\mb{a}_0+\epsilon^f_{\vec k+\vec{\mb{a}}}\right) f,
\eeq
where $f$ above is in its frequency-momentum representation, and $\epsilon^f$ is the mean-field dispersion of $f$. The action of $\mb{a}$, $S_{[\mb{a}]}=\int d\tau d^2x\mc{L}_{[\mb{a}]}$, is of the usual Maxwell-Yang-Mills form:
\beq
\mc{L}_{[\mb{a}]}=\frac{1}{2e^2}\tilde f_{\mu\nu}\tilde f_{\mu\nu}+\frac{1}{2g^2}\Tr\left(f_{\mu\nu}f_{\mu\nu}\right),
\eeq
where $e$ and $g$ are respectively the $U(1)$ and $SU(2)$ gauge couplings (not necessarily the same), $\tilde f_{\mu\nu}=\partial_\mu\tilde a_\nu-\partial_\nu\tilde a_\mu$ and $f_{\mu\nu}=\partial_\mu a_\nu-\partial_\nu a_\mu-i[a_\mu, a_\nu]$ are the $U(1)$ and $SU(2)$ field strengths, respectively. We have assumed that the speeds for the $U(1)$ and $SU(2)$ gauge fields are the same, and will show later that this causes no loss of generality as far as the universal low-energy physics is concerned.

In the above theory, when $B$ is condensed, the resulting state is a FL metal (or possibly more exotic variants thereof, as discussed in Sec. \ref{subsec: metals}). On the other hand, when $B$ is gapped, the resulting state can be a SRE insulator (see Sec. \ref{subsec: insulators}). For this theory to indeed describe a continuous DMT, the only relevant perturbation should be the one parameterized by $m^2$, the tuning parameter of this transition. We have already assumed (without any justification for now) that pairing of $f$ is relevant in a RG sense when $B$ is gapped. The important question that remains to be addressed is whether pairing is (ir)relevant at the critical point. Only if pairing is (dangerously) irrelevant at the critical point, can the above critical theory describe a continuous DMT.

A framework for analysing the critical theory of a simpler version of the above model has been developed in earlier works \cite{Senthil2008, Mross2010, Metlitski2014}. We summarize the key ideas borrowed from previous work and associated with the setup for studying the above transition here; see Ref.~\cite{Senthil2008} for a detailed discussion. We begin by focusing on the criticality associated with the $B$-sector while ignoring its coupling to $\mb{a}$ and $f$, which is expected to be described by a Wilson-Fisher type fixed point. However, due to the presence of the $f$-FS, $\mb{a}$ will be Landau-damped, which drives the coupling between $\mb{a}$ and $B$ in $S_{[B, \mb{a}]}$ to be irrelevant . Furthermore, if the scaling dimensions of gauge invariant operators made of $B$'s are larger than $3/2$, the coupling between $B$ and $f$ is also irrelevant. Then $f$ and $\mb{a}$ are dynamically decoupled from the $B$ sector. We now turn to the $f$ and $\mb{a}$ sector. Even though this sector is dynamically decoupled from the $B$ sector, the latter plays an important role at low energies. In particular, the $B$-sector changes the dynamical critical exponent associated with the $f$ and $\mb{a}$ sector from what is expected naively at the RPA-level, and can thereby modify the conclusion regarding pairing instabilities. Specifically, $\epsilon=1$ when $B$ is gapped (see Eqs. \eqref{eq: gauge fields action U(1) x U(1)} and \eqref{eq: gauge fields action U(2)} for the definition of $\epsilon$), but $\epsilon=0$ when $B$ is at its own transition described by a $2+1$ dimensional conformal field theory.

To summarize, in this subsection we have presented a possible mechanism  {\footnote{Based on certain assumptions, which we discuss at length in subsequent sections.}} for a continuous DMT, and the key ingredient is that the $f$-pairing is dangerously irrelevant at the above critical point. In passing, we remark that the purpose of considering a $U(2)$ gauge theory (rather than just an $SU(2)$ gauge theory) is to have the possibility that pairing can be irrelevant at the critical point, due to the $U(1)$ gauge field.

\subsection{Deconfined Metal-Metal Transition} \label{subsec: FT mechanism}

Building on the above mechanism for a continuous DMT, we can present a mechanism for a continuous \DMMT, which is described by $S_{{\rm DM^2T}}$ in Eq. \eqref{eq: FT theory schematic} (with $S_{\DMT}$ as defined in Sec. \ref{subsec: DMT mechanism}). Using the terminology introduced in Sec. \ref{sec: intro}, we will view the $c$-electrons in $S_{\DMT}$ as `hot' and the $d$-electrons as `cold'. Because of the strong correlations at the criticality, $B$ and $f$ can be very incoherent, which may potentially render the coupling in Eq. \eqref{eq: cold-hot coupling} irrelevant. Then the cold electrons are just spectators at the DMT associated with the hot electrons and we obtain a continuous \DMMT between two metals, where the size of the FS on either side of the transition (\ie in the two metallic phases) are different and finite. Moreover, there are only electronic FS in the two metallic phases and no neutral FS of any other emergent excitations.

As we have repeatedly emphasized, we have not examined some of our crucial assumptions behind the above mechanisms critically. One of our central assumptions has been related to the nature of the pairing instability associated with the $f$-fermions and the possibility of them being dangerously irrelevant at the critical point. In the next section, we will turn to a RG framework to examine these issues in a lot more critical detail.

\section{Renormalization-group analysis} \label{sec: RG}

This section will be devoted to a study of the $f$-FS coupled to a dynamical $U(2)$ gauge field, where $B$ is taken to be gapped. In addition to its potential relevance for the mechanism for a DMT as explained previously, it might also play an important role in the physics of Kitaev quantum spin liquids \cite{Zou2018}.

\subsection{$U(1)\times U(1)$ gauge theory} \label{subsec: 2-U(1) RG}

Before directly tackling the technically challenging problem of a FS coupled to a $U(2)$ gauge field, we first study a simpler version of the problem. Consider two $(2+1)$-dimensional FS coupled to two $U(1)$ gauge fields, $a_c$ and $a_s$. Both FS carry the same charge under $a_c$, but they carry opposite charges under $a_s$. This implies that $a_c$ ($a_s$) mediates repulsive (attractive) long-range Amperean interaction between fermions on antipodal regions of the FS, \ie $a_s$ favors pairing of the fermions while $a_c$ suppresses this tendency towards pairing \cite{Mross2010, Metlitski2014}. The fate of the FS at low-energies towards pairing thus depends on the competition between $a_c$ and $a_s$, which is a simpler version of the full problem of FS coupled to the $U(2)$ gauge field. Moreover, even within this simpler setting, arguments analogous to those in Sec. \ref{sec: U(2) gauge theory} can be used to construct interesting quantum phases. Furthermore, the introduction of an additional boson that carries opposite charges as the fermions under the above gauge fields can be used to study DMT and \DMMT.

We note that the problem with the two independent $U(1)$ gauge fields has appeared in a variety of different contexts in previous works, most notably in the study of the bilayer quantum Hall problem \cite{Bonesteel1996, Zou2016, Sodemann2016, Isobe2016}. However, in the context of bilayer quantum Hall systems, long-range Coulomb interactions play a defining role in the problem, while in our setting due to the charge gap they can be ignored in the low-energy regime of our interest. This distinction makes the physics in the two settings different. We will develop a RG framework that goes beyond the considerations of previous work. At the leading nontrivial order, we find that the result of the competition between $a_c$ and $a_s$ and the stability of the FS against pairing are determined by non-universal microscopic details. The extensions of these results beyond leading order will be the subject of future work; we will outline some of the interesting scenarios in Sec. \ref{sec: discussion}.

The action of the $U(1)\times U(1)$ problem is given by $S=\int d\tau d^2x \mc{L}$, where the Lagrangian has the form,
\beq
\mc{L}=\mc{L}_{[f_1, a_c+a_s]}+\mc{L}_{[f_2, a_c-a_s]}+\mc{L}_{[a_c]}+\mc{L}_{[a_s]}.
\label{2u(1)}
\eeq
The fermion fields, $f_1$ and $f_2$, carry charges under the even and odd combinations of the gauge fields, $a_c+a_s$ and $a_c-a_s$, respectively. We will also assume the following additional exchange symmetry: $f_1\leftrightarrow f_2$, $a_s\rightarrow - a_s$. We will now consider the case where $f_1$ and $f_2$ form Fermi surfaces, and our goal is to understand the infrared (IR) properties of this system.

\begin{figure}[h]
\begin{center}
\includegraphics[scale=0.22]{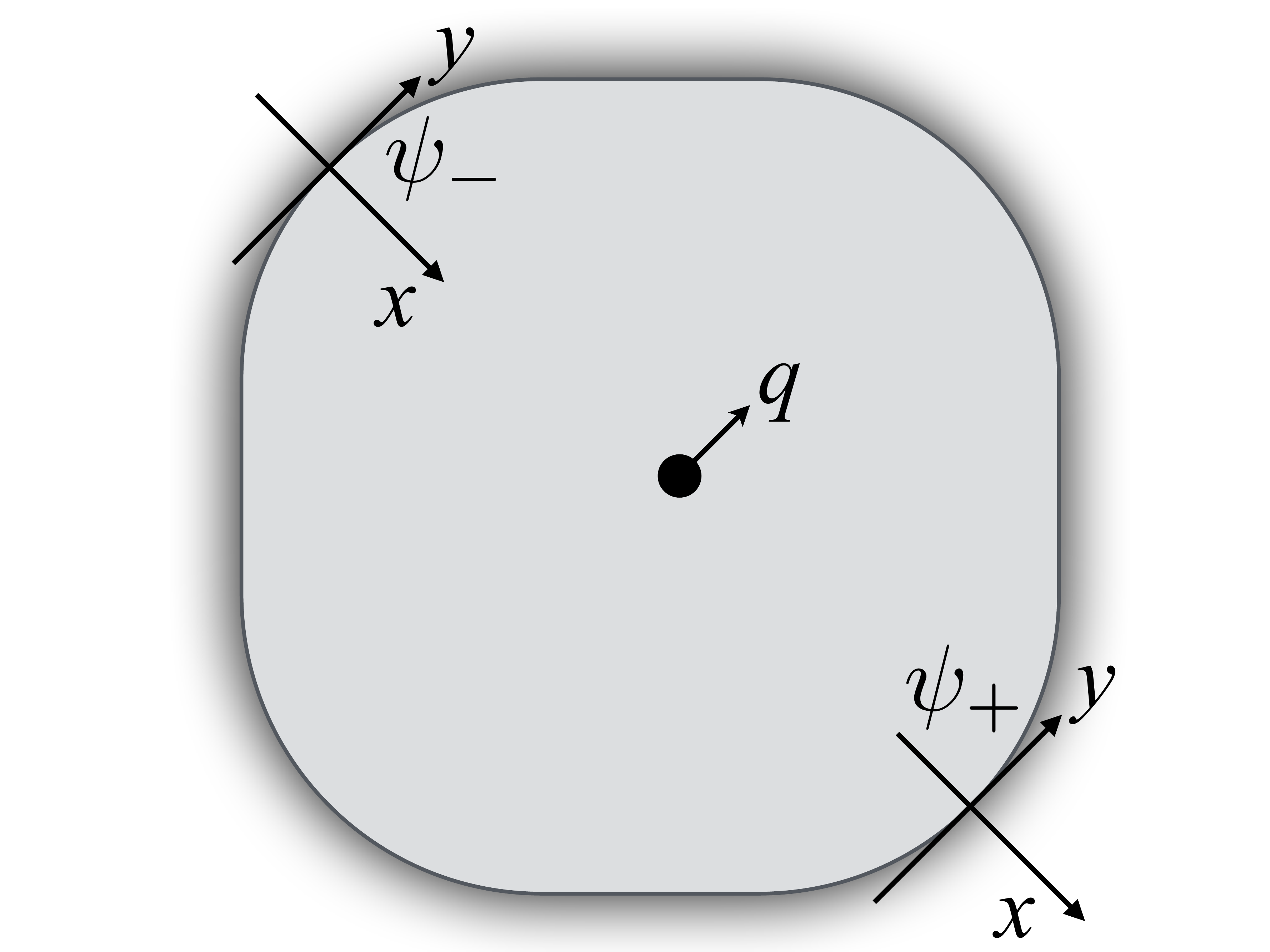}
\end{center}
\caption{A pair of antipodal patches on the FS that have nearly collinear Fermi velocities. The momentum of the gauge mode, $\vec q$, is nearly perpendicular to these Fermi velocities.}
\label{FS}
\end{figure}

As a result of minimal coupling to the fermion density, the temporal components of the gauge fields, $[a_c]_0$ and $[a_s]_0$, are screened out in a manner similar to the screening of Coulomb interactions in a metal, and they will be ignored in the following discussion. By working in the Coulomb gauge, $\vec\nabla\cdot\vec a_{c,s}=0$, we can focus on the transverse components of the gauge fields coupled minimally to the FS. Previous studies focusing on the problem of a single $U(1)$ gauge field coupled to a Fermi surface \cite{Lee2009,Metlitski2010,Mross2010} identified the importance of coupling of the small-momentum ($\vec q$) fluctuations of the transverse components of the gauge field to the low-energy excitations of the fermions near pairs of antipodal patches with nearly parallel Fermi velocities (see Fig.~\ref{FS}). We will adopt the same strategy below and consider a ``patch" description of the low-energy theory.

\subsubsection*{Patch theory: symmetries and scaling} \label{subsubsec: patch theory}

Let us introduce the fields, $\psi_{\alpha p}$, which denote the low-energy fermionic fields near the antipodal patches labeled by $p=\pm$ and color index, $\alpha=1,2$ (Fig.~\ref{FS}). Based on Refs. \cite{Lee2009, Metlitski2010, Mross2010}, the Lagrangian in an explicit form is given by
\begin{widetext}
\beq \label{eq: 2-U(1) patch}
\mc{L} = \mc{L}_f + \mc{L}_{[a_c, a_s]},
\eeq
with
\beq \label{eq: 2-U(1) patch fermion}
\begin{split}
\mc{L}_f=
\sum_{p=\pm,\alpha=1,2}\psi^\dagger_{\alpha p}\left[\eta\partial_\tau + v_F\left(-ip\partial_x -\partial_y^2\right)\right]\psi_{\alpha p}- v_F \sum_{p=\pm}\lambda_p\left[(a_c+a_s)\psi_{1p}^\dag \psi_{1p}+(a_c-a_s)\psi_{2p}^\dag \psi_{2p}\right]
\end{split}
\eeq
and
\beq \label{eq: gauge fields action U(1) x U(1)}
\mc{L}_{[a_c, a_s]}=\frac{N}{2e_c^2}|k_y|^{1+\epsilon}|a_c|^2+\frac{N}{2e_s^2}|k_y|^{1+\epsilon}|a_s|^2,
\eeq
\end{widetext}
where $v_F$ denotes the Fermi velocity, and $\lambda_\pm=\pm 1$. The time-derivative terms of the gauge fields are irrelevant \cite{Lee2009, Metlitski2010, Mross2010}, so they have been dropped off. Then the difference in the speed of ``photon" corresponding to $a_c$ and $a_s$ can be absorbed into the gauge couplings, $e_c$ and $e_s$, whose flows will be the subject of subsequent discussion. Notice that we have introduced two additional parameters --- $N$, the number of flavors of fermions $f_{1,2}$, and $\epsilon$ --- with an eye for doing a controlled double-expansion in small $\epsilon$ and large $N$ \cite{Nayak1993, Mross2010}. Notice, in principle, that we can choose the $\epsilon$'s for $a_c$ and $a_s$ to be different, but we will take them to have the same value in accord with the physical case, where, for example, $\epsilon=1$ corresponds to the usual example of a spinon Fermi surface coupled to a $U(1)$ gauge field \cite{Lee1992}, and, $\epsilon=0$ corresponds to the $\nu=1/2$ Landau level problem for the composite Fermi liquid with Coulomb interactions \cite{Halperin1993}. The physical case we focus on corresponds to $\epsilon=1$.

A few general remarks are in order here. 
\begin{itemize}

    \item Starting with the above low-energy field theory, Eq. \eqref{eq: 2-U(1) patch}, we can either (i) fix $v_F=1$ and study the flow of $\eta$, or, (ii) fix $\eta=1$ and study the flow of $v_F$, in addition to the flow of the coupling constants $e_c^2,~e_s^2$. It is easy to see that we can go back and forth between the two schemes by appropriate rescalings of the $\psi$-fields. For the remainder of our discussion, we will adopt scheme (i) above. In this case, the physical Fermi velocity is $1/\eta$.
    
    \item The above Lagrangian only describes a particular pair of antipodal patches on the Fermi surfaces. In order to describe the whole system, we need to include all the patches defined along the entire contour of the Fermi surface; we will consider the effects of inter-patch interactions below after analyzing the scaling structure and key features associated with a single pair of patches. 
    
    \item Let us review the general structure of the theory defined in Eqn.~\eqref{eq: 2-U(1) patch}. As noted in Ref.~\cite{Metlitski2010}, the theory has an emergent rotational symmetry:
\beq
\begin{split}
&a_{c,s}(\tau, x,y)\rightarrow a_{c,s}(\tau, x, y+\theta x)\\
&\psi_{\alpha p}(\tau, x, y)\rightarrow e^{-ip\left(\frac{\theta}{2}y+\frac{\theta^2}{4}x\right)}\psi_{\alpha p}(\tau, x, y+\theta x)
\end{split}
\eeq	
which equivalently in momentum space reads
%\begin{widetext}
\beq \label{rotsym}
%\begin{split}
&a_{c,s}(\omega, q_x, q_y)\rightarrow a_{c,s}(\omega, q_x-\theta q_y, q_y)\nonumber\\
&\psi_{\alpha p}(\omega, q_x, q_y)\rightarrow \psi_{\alpha p}\left(\omega, q_x-\theta q_y-p\frac{\theta^2}{4}, q_y+p\frac{\theta}{2}\right).\nonumber\\
%\end{split}
\eeq
%\end{widetext}
A direct consequence of the above rotational symmetry is that the propagators satisfy the following properties:
\beq
\begin{split}
&D_{c,s}(\omega, q_x, q_y)= D_{c,s}(\omega, q_y)\\
&G_{\alpha p}(\omega, q_x, q_y)=G_{\alpha p}(\omega, pq_x+q_y^2),
\end{split}
\eeq
\ie the gauge field propagator does not depend on $q_x$, and the Fermi surface curvature does not flow under RG. (In more technical terms, $ip\psi^\dag\partial_x\psi$ and $\psi^\dag\partial_y^2\psi$ renormalize in the same way). 

Because of the forms of these propagators, the relation between the dynamical exponents of $a_c$, $a_s$ and $\psi$, denoted $z_{a_{c,s}}$ and $z_{f}$,  respectively, are also fixed: $z_{a_c}=z_{a_s}=2z_{f}$. Notice here $z_{a_{c,s}}$ is the relative scaling between $\omega$ and $q_y$ in the gauge field propagators, while $z_f$ is the relative scaling between $\omega$ and $q_x$ in the fermion propagator.

\item When $\epsilon<1$, due to the non-analytic nature of the kinetic terms of the gauge fields, $z_{a_c}=z_{a_s}=2+\epsilon$ exactly. This implies a great simplification in the discussion of pairing instability, as discussed below. However, because $\epsilon=1$ in the absence of additional gapless bosons that carry opposite charges as the fermions under $a_c$ and $a_s$, we cannot directly utilize the properties associated with the case with $\epsilon<1$. Nevertheless, the case with $\epsilon=0$ is pertinent to the critical points of DMT and DM$^2$T, which will be discussed briefly at the end of this subsection.

\item Consider now the following scaling transformations for the frequency and momenta,
\beq \label{eq: barescal}
\omega\rightarrow e^{z_f\l}~\omega,~~q_x\rightarrow e^{\l}~q_x,~~q_y\rightarrow e^{\l/2}~q_y,
\eeq
which leaves the bare kinetic energy for $\psi_{\alpha p}$ in Eq.~\eqref{eq: 2-U(1) patch} invariant with $z_f=1$ (at tree-level). Under the above scaling, the fields transform as
\beq
\psi_{\alpha p}\rightarrow e^{\left(\frac{1}{4}+\frac{z_f}{2}\right)\l}~ \psi_{\alpha p},~a_{c,s}\rightarrow e^{\l}~a_{c,s},
\eeq
and by simple power counting, in order to make the whole Lagrangian, Eq. \eqref{eq: 2-U(1) patch}, invariant, the scaling transformation for $e_{c,s}$ should be
\beq
e_{c,s}^2\rightarrow e^{\left(1+\frac{\epsilon}{2}-z_f\right)\l}~e_{c,s}^2.
\label{tree}
\eeq
Because $z_f=1$ at the tree level, the above scaling indicates that both gauge couplings are relevant as long as $\epsilon>0$, which is the case of our interest.

\end{itemize}

\subsubsection*{RG analysis and pairing instability} \label{subsubsec: pairing instability}

In order to study the low-energy properties of this two-patch theory, we now apply an RG analysis to go beyond the tree level scaling considerations. In Appendix \ref{app: Wilsonian RG}, we develop a framework for Wilsonian RG that can be used to calculate the flows of these coupling constants systematically. Our detailed derivation also clarifies a number of conceptual issues that are potentially beneficial for the readers. We apply this framework to calculate the beta functions of the couplings in Appendix \ref{app: fermion propagator}.

We focus on dimensionless gauge coupling constants,
\beq
\alpha_c=\frac{e_c^2}{4\pi^2\eta\Lambda^\epsilon},
\quad
\alpha_s=\frac{e_s^2}{4\pi^2\eta\Lambda^\epsilon}.
\eeq
where $\Lambda$ is the cutoff on $q_y$. Their beta functions (defined explicitly in Appendix \ref{app: Wilsonian RG}) to the next leading order are {\footnote{To this order, these beta functions can be viewed as a result of either an expansion in terms of small $\epsilon$ but finite $N$, or a double expansion of small $\epsilon$ and large $N$.}}
\beq \label{eq: beta function dimensionless 2-U(1)}
\begin{split}
&\beta(\alpha_c)=\left(\frac{\epsilon}{2}-\frac{\alpha_c+\alpha_s}{N}\right)\alpha_c\\
&\beta(\alpha_s)=\left(\frac{\epsilon}{2}-\frac{\alpha_c+\alpha_s}{N}\right)\alpha_s.
\end{split}
\eeq

From these beta functions, we get a nontrivial fixed {\it line} at:
\beq \label{eq: fixed line 2-U(1)}
\alpha_c+\alpha_s=\frac{\epsilon N}{2}.
\eeq
We also notice that the above beta functions imply that
\beq \label{eq: RG invariance 2-U(1)}
\beta\left(\frac{e_c^2}{e_s^2}\right)=\beta\left(\frac{\alpha_c}{\alpha_s}\right)=0
\eeq
That is, the ratio of the gauge couplings, $e_c^2/e_s^2=\alpha_c/\alpha_s=r$, is RG invariant to this order. Therefore, the beta functions at this order predict that in the IR the fixed point values for $\alpha_c$ and $\alpha_s$ are given by
\beq \label{eq: leading fixed point 2-U(1)}
\begin{split}
&\alpha_{c*}=\frac{\epsilon N}{2}\cdot\frac{r}{r+1}=\frac{r}{2(r+1)},\\
\quad
&\alpha_{s*}=\frac{\epsilon N}{2}\cdot\frac{1}{r+1}=\frac{1}{2(r+1)},
\end{split}
\eeq
which is determined by the non-universal ratio, $r$. Notice that the physical values of $N=1$ and $\epsilon=1$ have been substituted in the last step.

\begin{figure}[h]
\begin{center}
\includegraphics[scale=0.5]{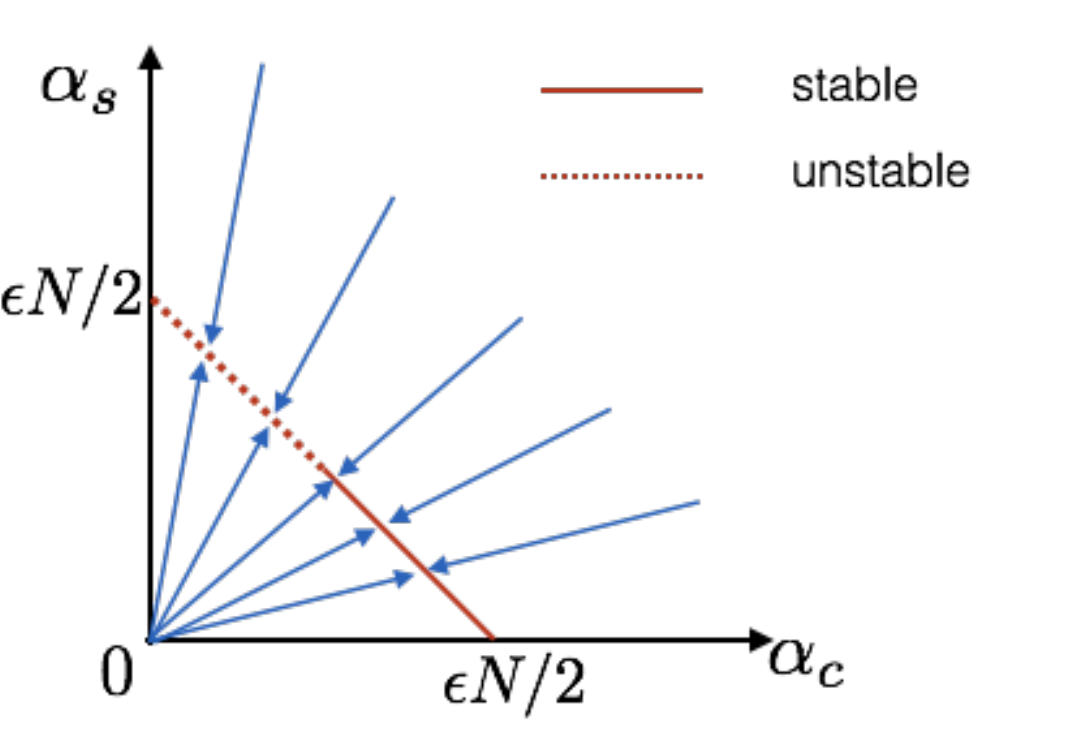}
\end{center}
\caption{The leading-order RG flow of $\alpha_c$ and $\alpha_s$ in the $U(1)\times U(1)$ gauge theory, where the blue arrows represent the direction of the flow. The red line represents the nontrivial fixed line, where the solid (dashed) part represents the regime stable against (unstable to) pairing. Replacing $\alpha_c$ by $\tilde\alpha$ and $\alpha_s$ by $\alpha$, this figure represents the leading-order RG flow of the $U(2)$ gauge theory.}
\label{fig: flow}
\end{figure}

Now we use the above results to analyze the pairing instability, via the approach in Ref. \cite{Metlitski2014}.  We fix $N=1$ while still take $\epsilon$ to be a small expansion parameter. We focus on the RG flow of the dimensionless four-fermion interactions in the BCS $s$-wave color-singlet channel, denoted by $\widetilde V$ (see Appendix \ref{app: BCS coupling} for a precise definition of this dimensionless coupling constant). Results for other channels can be obtained similarly, and we ignore them here for simplicity.

There are two contributions to the flow of $\widetilde V$. The first is the usual FL contribution \cite{SHANKAR1991, Polchinski1992, Shankar1993, Metlitski2014}:
\beq
\left.\beta(\widetilde V)\right|_{\rm FL}=-\widetilde V^2.
\eeq
The second contribution comes from the coupling between the fermions and the gauge fields. In particular, as already pointed out in previous work \cite{Metlitski2014}, integrating out the fast gauge modes along with shrinking and breaking each patch of the FS generates inter-patch interactions. Generalizing these calculations to our setting in a straightforward fashion, we find that at leading order, the gauge fields contribute in the following way to the flow of $\widetilde V$,
\beq
\left.\beta(\widetilde V)\right|_{\rm gauge}=\alpha_c-\alpha_s.
\eeq
As expected, the first (second) term arises from the minimal coupling to $a_c$ ($a_s$), which suppresses (enhances) pairing. Taken together, the full beta function at this order is given by,
\beq
\beta(\widetilde V)=\alpha_c-\alpha_s-\widetilde V^2.
\eeq

Combining the above flow equation for $\widetilde{V}$ with Eqs. \eqref{eq: beta function dimensionless 2-U(1)}-\eqref{eq: leading fixed point 2-U(1)}, we see that at the leading nontrivial order, the propensity of the FS towards pairing depends on non-universal microscopic details, parametrized by $r$ (introduced around Eq. \eqref{eq: leading fixed point 2-U(1)}). More precisely, if $r<1$, $a_s$ wins in the competition, $\widetilde V$ flows to $-\infty$ in the IR and the FS are unstable to pairing. On the other hand, if $r>1$, $a_c$ wins in the competition, as long as the UV value of $\left.\widetilde V\right|_{\rm UV}>-\sqrt{\frac{r-1}{2(r+1)}}$. In the IR $\widetilde V$ flows to $\widetilde V_*=\sqrt{\frac{r-1}{2(r+1)}}>0$, and the FS are perturbatively stable against pairing. If $r=1$, the effects of the two gauge fields exactly cancel each other and $\widetilde V$ flows to $-\infty$ if it starts with a negative value in the UV, in which case the FS are unstable. If $\widetilde V$ starts with a positive value in the UV, it flows to 0, and the FS are stable. {\footnote{Until the Kohn-Luttinger mechanism for pairing eventually becomes important \cite{Kohn1965, Shankar1993}, which we will ignore throughout this paper.}} These results are summarized in the flow diagram in Fig. \ref{fig: flow}.

When higher order corrections to the beta functions in the two-patch theory are included, the fixed line described by Eq. \eqref{eq: fixed line 2-U(1)} might collapse into interesting fixed points. Furthermore, in order to be internally self-consistent, including higher order corrections to $\beta(\alpha_c)$ and $\beta(\alpha_s)$ requires one to include the corrections to $\beta(\widetilde V)$ to the same order. While clearly interesting, this is beyond the scope of the present work; we will instead discuss the possible scenarios in Sec. \ref{sec: discussion} and leave a detailed systematic study of these corrections to future work.

\subsubsection*{Possible DMT and DM$^2$T critical points}

Before finishing the discussion on this $U(1)\times U(1)$ gauge theory, we note that an interesting simplification occurs when $\epsilon=0$, which may be relevant for DMT and DM$^2$T critical points. As discussed in Appendix \ref{app: Wilsonian RG}, due to the non-analyticity of the kinetic terms of the gauge fields and the associated absence of an anomalous dimension, the ratio between the two gauge couplings is exactly RG invariant when $\epsilon=0$ (not just to the leading nontrivial order). In this case, it is sufficient to only consider the beta functions at the leading nontrivial order without worrying about higher order corrections. These beta functions are given by
\beq \label{eq: beta functions epsilon=0 U(1) x U(1)}
\begin{split}
    &\beta(\alpha_c)=-(\alpha_c+\alpha_s)\cdot\alpha_c\\
    &\beta(\alpha_s)=-(\alpha_c+\alpha_s)\cdot\alpha_s\\
    &\beta(\widetilde V)=\alpha_c-\alpha_s-\widetilde V^2
\end{split}
\eeq
The flows of these couplings are discussed in detail in Appendix \ref{app: RG U(1) x U(1) epsilon=0}. It is shown that when $r<1$, the FS is unstable, while the FS is perturbatively stable when $r\geqslant 1$. Notice when $r=1$, one has to go to the next leading order of the third equation above to analyze the effects of gauge fluctuations on pairing, as discussed in Appendix \ref{app: RG U(1) x U(1) epsilon=0}.

This result is useful for constructing a DMT and DM$^2$T critical point, because if additional bosons with opposite charges as the fermions under $a_c$ and $a_s$ are introduced, the mechanism discussed in Sec. \ref{sec: transitions} can also be applied to this $U(1)\times U(1)$ gauge theory to describe a DMT and DM$^2$T. Because these additional bosons are the analog of the $B$ bosons discussed in Sec. \ref{sec: transitions}, here we also refer to them as $B$ bosons. Recall that in this mechanism, the DMT and DM$^2$T are driven by the condensation transition of these $B$ bosons, and it is important that pairing is irrelevant at the transition, in order for these transitions to be continuous without fine-tuning. When the condensation transition of the $B$ bosons is described by a $2+1$ dimensional conformal field theory, Ref. \cite{Senthil2008} pointed out that $\epsilon=0$ right at the transition. From our discussion above, we see that pairing can indeed be irrelevant at the transition where $\epsilon=0$. This observation provides more evidence that this theory can potentially describe DMT and DM$^2$T, if one can further show that pairing becomes relevant immediately after $B$ becomes gapped. Suppose this pairing instability indeed exists when $B$ is gapped, then one concrete example of DMT can be realized in a bilayer system, where each layer undergoes the bandwidth-tuned Mott transition described in Ref. \cite{Senthil2008}, and the two layers are weakly coupled in a way that preserves a layer-exchange symmetry. In fact, as shown in Ref. \cite{Zou2016}, in this case right at the DMT the two layers are effective decoupled because interlayer couplings are irrelevant, and the critical point hosts sharply defined FS without any long-lived quasiparticles.

\subsection{$U(2)$ gauge theory} \label{subsec: U(2) RG}

We now return to the problem of FS coupled to a $U(2)$ gauge field. Compared to the $U(1)\times U(1)$ problem, there are now potential complications due to the non-Abelian nature of this gauge group. We will show that many of the anticipated complications do not play an essential role as a result of the underlying FS, and the problem reduces to a form similar to the $U(1)\times U(1)$ problem studied above.

The action now has the seemingly more complicated form,
\beq \label{eq: U(2) action UV}
S=S_{[f,\mb{a}]}+S_{[\mb{a}, \Phi]}+S_{\rm FP},
\eeq
where $\mb{a}=a+\tilde a \mb{1}$ is a $U(2)$ gauge field, with $a$ its $SU(2)$ part and $\tilde a$ its $U(1)$ part. The field $f$ is the fermion field that at finite density forms FS. The auxiliary field, $\Phi$, and $S_{\rm FP}$, corresponding to the action for the Faddeev-Popov (FP) ghosts, are introduced for the sake of gauge fixing in this non-Abelian gauge theory \cite{Peskin1995}. As before, we will use the Coulomb gauge, \ie $\vec\nabla\cdot\vec{\mb{a}}=0$, where the action has the explicit form, 
\begin{widetext}
\beq
S_{[f, \mb{a}]}=\int_{\vec x, \tau}f_{{\vec x},\tau}^\dag(\partial_\tau-\mu_f+i\mb{a}_0)f_{{\vec x},\tau} + \int_{\vec k,\omega}f^\dag_{\vec k,\omega}\epsilon^f_{\vec k+\vec{\mb{a}}}f_{\vec k, \omega},
\eeq
\beq
S_{[\mb{a}, \Phi]}=\int_{\vec x, \tau}\left(\frac{1}{2e^2}\tilde f_{\mu\nu}^2+\frac{1}{4g^2}\left(f_{\mu\nu}^\alpha\right)^2+\frac{\xi}{2}\left(\Phi^\alpha\right)^2+\Phi^\alpha\vec\nabla\cdot\vec{a}^\alpha\right),
\eeq
\end{widetext}
and
\beq \label{eq: FP action}
S_{\rm FP}=\int_{\vec x, \tau}\bar c^\alpha \left(-\vec\nabla\cdot\vec D^{\alpha\beta}_{\rm adj}\right)c^\beta,
\eeq
where the field strengths are $\tilde f_{\mu\nu}=\partial_\mu\tilde a_\nu-\partial_\nu\tilde a_\mu$ and $f_{\mu\nu}^\alpha=\partial_\mu a^\alpha_\nu-\partial_\nu a^\alpha_\mu+\epsilon^{\alpha\beta\gamma}a^\beta_{\mu}a^\gamma_{\nu}$, with $\epsilon^{\alpha\beta\gamma}$ the fully anti-symmetric tensor satisfying $\epsilon^{123}=1$. We choose the generators of $SU(2)$ in the fundamental representation to be $T^\alpha=\sigma^\alpha/2$, with $\sigma^\alpha$ the standard Pauli matrices, such that $\left[T^\alpha, T^\beta\right]=i\epsilon^{\alpha\beta\gamma}T^\gamma$. The dispersion $\epsilon^f$ is taken to exhibit FS at the mean-field level. The covariant derivative in the adjoint representation is $D^{\alpha\beta}_{{\rm adj},\mu}\equiv\partial_\mu\delta^{\alpha\beta}-i\left(T^\gamma_{\rm adj}\right)^{\alpha\beta}a_\mu^\gamma\equiv\partial_\mu\delta^{\alpha\beta}-\epsilon^{\alpha\beta\gamma}a^\gamma_\mu$. The Grassman fields, $c$ and $\bar c$ are the Faddeev-Popov ghost and anti-ghost, respectively. The field $\Phi^\alpha$ is an auxiliary field to implement the gauge fixing, and the Coulomb gauge is achieved in the limit $\xi\rightarrow 0$.

As we can see, the present $U(2)$ gauge theory is very similar to the previous $U_c(1)\times U_s(1)$ problem, where the the $SU(2)$ gauge field $a$ plays a role analogous to $a_s$ in the latter problem. One therefore expects that the stability of the FS will be determined by the competition between the $U(1)$ gauge field, $\tilde a$, and the $SU(2)$ gauge field, $a$, where the former tends to suppress pairing and hence stabilize the FS, while the latter tends to enhance the tendency towards pairing.

There are a few notable differences between the two setups, which we now highlight. The first difference arises from the self-interaction of $a$ due to its non-Abelian nature. However, in the low-energy limit we will see that within the patch formulation the self-interaction of $a$ becomes irrelevant in an RG sense \cite{Sachdev2009}. Another notable difference for the $U(2)$ problem is related to the presence of the Faddeev-Popov ghosts. In the literature on color superconductivity, the ghosts are often argued to be unimportant using the ``hard-dense loop approximation" \cite{Shovkovy2004, Alford2007}. We will also show that the ghosts become irrelevant for the low-energy physics within the patch formulation. Thus, in spite of these differences, ultimately the $U(2)$ gauge theory coupled to FS becomes similar to the $U_c(1)\times U_s(1)$ gauge theory coupled to FS.

Before moving to a patch formulation of the theory, it is useful to examine the remaining gauge symmetry and the BRST symmetry of this theory \cite{Peskin1995}. In the Coulomb gauge, there is a remaining gauge symmetry:
\beq
\begin{split}
	&f\rightarrow Uf\\
	&\tilde a_\mu\rightarrow \tilde a_\mu-\partial_\mu\tilde\theta(t)\\
	&a_\mu\rightarrow Ua_\mu U^\dag+iU\partial_\mu U^\dag,
\end{split}
\eeq
with $U=e^{i\left(\tilde\theta(t)+\theta_\alpha(t)T^\alpha\right)}$. Notice that the $\theta$'s depend only on time but not on space. In its infinitesimal form, the BRST symmetry reads,
\beq
\begin{split}
	&\delta a_\mu^\alpha=\theta_B D^{\alpha\beta}_{{\rm adj}, \mu}c^\beta\\
	&\delta f=i\theta_B c^\alpha T^\alpha f\\
	&\delta c^\alpha=-\frac{1}{2}\theta_B \epsilon^{\alpha\beta\gamma}c^\beta c^\gamma\\
	&\delta\bar c^\alpha=\theta_B \Phi^\alpha\\
	&\delta \Phi^\alpha=0
\end{split}
\eeq
with $\theta_B$ an infinitesimal Grassmann variable. Constrained by these symmetries among with other symmetries in the system, we are able to write down the most general effective action. In particular, we notice that no other bilinears of the ghost and anti-ghost fields with less than two spatial derivatives can be written down (\eg a ``chemical potential" type term $\bar c^\alpha c^\alpha$ is forbidden). 
\\

\subsubsection*{Patch theory} \label{subsubsec: patch U(2)}

As before, due to the finite density of $f$ fermions, the temporal components of the gauge field are screened and we ignore them in the following treatment. Moreover, the Coulomb gauge allows us to keep only the transverse components of the spatial components of the gauge fields. Expanding the fermion operators in the vicinity of the Fermi surface and focusing only on pairs of antipodal regions with nearly parallel Fermi velocities, we obtain the patch formulation of the theory Eq. \eqref{eq: U(2) action UV}:
\beq
\mc{L}=\mc{L}_{[\psi, \mb{a}]}+\mc{L}_{[\mb{a}]}+\mc{L}_{{\rm irre}},
\eeq
with
\begin{widetext}
\beq
\mc{L}_{[\psi, \mb{a}]}=\sum_{p=\pm}\psi^\dag_{p}\left[\eta\partial_\tau+\left(-ip\partial_x-\partial_y^2\right)\right]\psi_p-\sum_{p=\pm}\lambda_p\psi^\dag_{[p\alpha]}\mb{a}_{\alpha\beta}\psi_{[p\beta]},
\eeq
\end{widetext}
and
\beq \label{eq: gauge fields action U(2)}
\mc{L}_{[\mb{a}]}=\frac{N}{2e^2}|k_y|^{1+\epsilon}|\tilde a|^2+\frac{N}{4g^2}|k_y|^{1+\epsilon}|a^\alpha|^2,
\eeq
where $\psi$ is the low-energy fermionic operator in the patch formulation, and $\mb{a}$ now only represents the transverse components of the spatial components of the gauge field. Notice we have already introduced $N$ and $\epsilon$, just as in the $U_c(1)\times U_s(1)$ case, with an eye for doing a controlled double-expansion. 

A useful starting point, as before, is to consider the scaling structure of the patch theory:
\beq
\begin{split}
&\omega\rightarrow e^{z_f\ell}\omega,
\quad
k_x\rightarrow e^\ell k_x,
\quad
k_y\rightarrow e^{\frac{\ell}{2}}k_y,\\
&\psi(\vec x, \tau)\rightarrow e^{\left(\frac{1}{4}+\frac{z_f}{2}\right)\ell}\psi(\vec x, \tau),\\
&\mb{a}(\vec x, \tau)\rightarrow e^{\ell}\mb{a}(\vec x, \tau)\\
&e^2\rightarrow e^{\left(1+\frac{\epsilon}{2}-z_f\right)}e^2,
\quad
g^2\rightarrow e^{\left(1+\frac{\epsilon}{2}-z_f\right)}g^2
\end{split}
\eeq
with $z_f=1$ at the tree level. Notice that according to the action of the ghosts, Eq. \eqref{eq: FP action}, the ghosts should transform as $c\rightarrow e^{3\ell/4} c$ under the above scaling.

From the above scaling structure, it is straightforward to verify that the self-interaction of $a$ and the coupling between the ghosts with the fermion-gauge sector are irrelevant. For example, consider the 3-gluon self-interaction, which schematically has the form $k_ya^3$. This term has a larger scaling dimension compared to $|k_y|^{1+\epsilon}|a^{\alpha}|^2$, so it is irrelevant. As another example, consider the coupling between the ghosts and the gauge field, which schematically has the form $k_ya\bar cc$. This coupling has a larger scaling dimension compared to $v_Fa\psi^\dag\psi$, so it is also irrelevant. We include all of the irrelevant terms in $\mc{L}_{{\rm irre}}$, which will be ignored in the following discussion. 

The main lesson we draw from the above scaling analysis is that as a result of the scaling structure that is tied to the presence of the underlying $f$-FS, the $SU(2)$ gauge field in the current $U(2)$ problem is effectively ``quasi-Abelianized". The problem thus effectively reduces to the previous setup involving $U_c(1)\times U_s(1)$ gauge fields. In particular, the features of the patch theory discussed in Sec. \ref{subsubsec: patch theory} and the RG framework developed in Appendix \ref{app: Wilsonian RG} can be directly applied here, once we (schematically) replace $a_c$ by $\tilde a$ and $a_s$ by $a$. The differences between the two problems in the low-energy limit arise mainly from the difference in the structure of the interaction vertices.

\subsubsection*{RG analysis and pairing instability} \label{subsubsec: RG U(2)}

Now we apply the RG framework developed in Appendix \ref{app: Wilsonian RG} to the $U(2)$ problem with appropriate coupling constants and interaction vertices. Our calculations will be performed to the leading order in $\mO(\epsilon)\sim O(1/N)$ (see Appendix \ref{app: fermion propagator} for details).

Focusing on the dimensionless gauge couplings,
\beq
\tilde\alpha=\frac{e^2}{4\pi^2\eta\Lambda^\epsilon},
\quad
\alpha=\frac{3g^2}{8\pi^2\eta\Lambda^\epsilon},
\eeq
the beta functions are given by {\footnote{As in the $U(1)\times U(1)$ gauge theory, to this order these beta functions can be viewed as the result of either an expansion of small $\epsilon$ but finite $N$, or a double expansion of small $\epsilon$ and large $N$.}}
\beq
\begin{split}
&\beta(\tilde\alpha)=\left(\frac{\epsilon}{2}-\frac{\tilde\alpha+\alpha}{N}\right)\tilde\alpha,\\
&\beta(\alpha)=\left(\frac{\epsilon}{2}-\frac{\tilde\alpha+\alpha}{N}\right)\alpha.
\end{split}
\eeq
We find once again that at this order there is a nontrivial fixed {\em line} described by
\beq
\tilde\alpha+\alpha=\frac{\epsilon N}{2}.
\eeq
The ratio between the gauge couplings is RG invariant, \ie $\beta\left(\tilde\alpha/\alpha\right)=0$. The values of $\tilde\alpha$ and $\alpha$ in the IR are given by,
\beq
\begin{split}
&\tilde\alpha_*=\frac{\epsilon N}{2}\cdot\frac{r}{r+1}=\frac{r}{2(r+1)},\\
&\alpha_*=\frac{\epsilon N}{2}\cdot\frac{1}{r+1}=\frac{r}{2(r-1)}
\end{split}
\eeq
where $r$ is the ratio $\tilde\alpha/\alpha$ at the cutoff scale, which is determined by non-universal microscopic details. Again, the physical values of $N=1$ and $\epsilon=1$ have been substituted in the last step.

Just like in the case of the $U_c(1)\times U_s(1)$ problem, we can now study the flow of the dimensionless BCS coupling in the $s$-wave color-singlet channel \cite{Metlitski2014} in the current setting, which at leading order of $\epsilon$ for $N=1$ is given by
\beq
\beta(\widetilde V)=\tilde\alpha-\alpha-\widetilde V^2.
\eeq
Not surprisingly, we find the nature of the pairing instability in the current setting to be similar to the $U_c(1)\times U_s(1)$ problem: the FS are stable against pairing depending on non-universal microscopic details parametrized by $r$. If $r>1$, the $U(1)$ gauge field wins in the competition and the FS are perturbatively stable. If on the other hand $r<1$, the $SU(2)$ gauge field wins in the competition and the FS are unstable to pairing. If $r=1$, the FS are stable against (unstable to) pairing if the microscopic short-range interactions among the fermions are repulsive (attractive). Once again, these results for the RG flows can be depicted schematically as in Fig. \ref{fig: flow}.

We emphasize here once again that  higher order corrections to the above beta functions can potentially cause the fixed line of $(\tilde\alpha, \alpha)$ to collapse into (undetermined) fixed points, which can have an effect on the nature of the pairing instability. These considerations are beyond the scope of the present work but we discuss some of the interesting scenarios in the final Sec. \ref{sec: discussion}, leaving a detailed analysis for future work.

We also stress that the argument leading to the simplification in the discussion of the pairing instability in the $U(1)\times U(1)$ gauge theory with $\epsilon=0$ does not directly apply to the $U(2)$ gauge theory, because the $SU(2)$ gauge field may acquire a nonzero anomalous dimension (see Appendix \ref{app: Wilsonian RG}).

\section{Discussion} \label{sec: discussion}

In this paper, we have shown that a $U(2)$ gauge theory based approach can describe a myriad of quantum phases, including conventional phases (\eg conventional insulators and FL metals with/without broken symmetries) and more exotic phases (\eg quantum spin liquids, orthogonal metals and fractionalized Fermi liquids). We have addressed the possible routes towards describing interesting quantum phase transitions between these phases, which include the deconfined Mott transition and Fermi (or, deconfined metal-metal) transition.

Using a renormalization group formalism, we have outlined and identified possible mechanisms (at leading order in our expansion parameters) through which such transitions can occur. In particular, we have noticed the existence of fixed lines (instead of the more familiar fixed points) at low energies in the two-patch theories, where based on the UV values of the gauge coupling constants, part of the line remains stable against pairing while the complementary part becomes unstable to pairing. It is important to remain cautious as corrections that are higher order in the expansion parameters can potentially alter the true infrared properties of these field theories. 

We have also studied a $U(1)\times U(1)$ gauge theory that shares many common aspects with the $U(2)$ gauge theory. This $U(1)\times U(1)$ gauge theory has many important implication on various phenomena. For example, if long-range interactions are included (as opposed to purely local interactions considered in this paper), we are able to propose a concrete setup where a continuous DMT can arise, and we can describe the universal physics associated with that DMT in a controlled manner \cite{Zou2020}. This $U(1)\times U(1)$ gauge theory is also directly related to the possible ground state of the intensely studied QSL candidates, including $\kappa$-(ET)$_2$Cu$_2$(CN)$_3$ and EtMe$_3$Sb[Pd(dmit)$_2$]$_2$ \cite{Shimizu2003, Yamashita2010}. These are layered materials, and are argued to host QSL with emergent neutral Fermi surfaces. Our results on the $U(1)\times U(1)$ gauge theory clearly raise concern on the stability of such QSL states as a true ground state in these layered materials.

\begin{figure}[h]
\begin{center}
\includegraphics[scale=0.45]{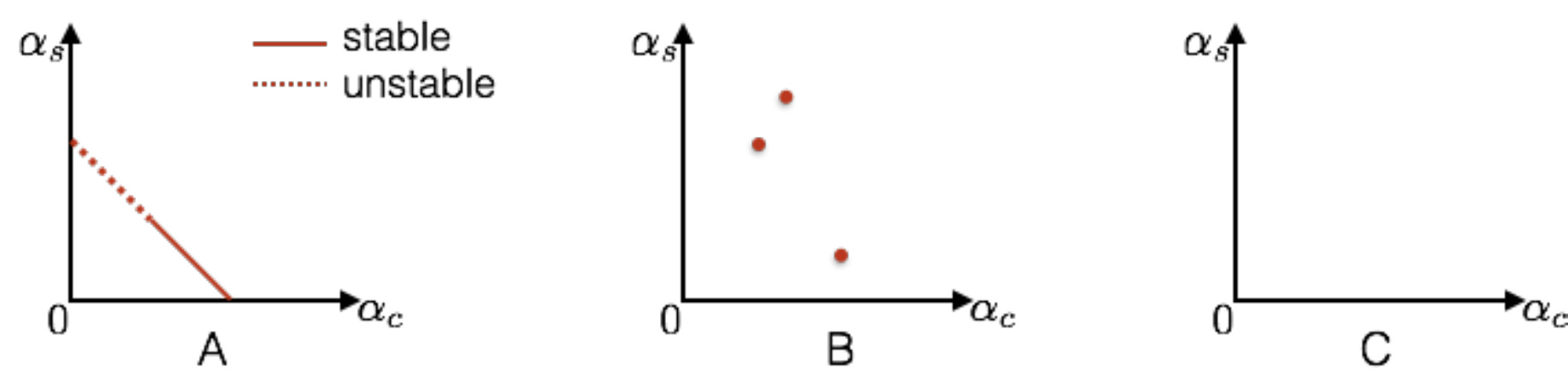}
\end{center}
\caption{Different scenarios of the $U(1)\times U(1)$ gauge theory after taking into account higher-order contributions. Notice in Scenario A the shape and position of the fixed line can potentially be modified compared to the leading-order result. Replacing $\alpha_c$ by $\tilde\alpha$ and $\alpha_s$ by $\alpha$ gives rise to the analogous scenarios for the $U(2)$ gauge theory.}
\label{fig: scenarios}
\end{figure}

We end this paper by discussing possible scenarios that may arise after higher order corrections are taken into account in the presence of only local interactions, and leave a detailed analysis of their effects for the future. All of these scenarios are depicted pictorially in Fig. \ref{fig: scenarios}.

\begin{enumerate}
    
    \item \underline{Scenario A}: The fixed lines for the two-patch theories discussed in Sec.  \ref{sec: RG} remain robust upon the inclusion of higher order corrections. If so, on part of the fixed line the FS remains stable against pairing. Then we can have an interesting scenario where two dimensional translationally invariant critical systems are described by fixed lines. On the fixed line, the system can be viewed as FS coupled to two dynamical $U(1)$ gauge fields or a dynamical $U(2)$ gauge field, which is a new type of QSL. Furthermore, these act as parent states for other interesting phases, obtained by tuning different parameters in the system (see Fig. \ref{fig: landscape}). However, in this case, it may be difficult to describe a continuous DMT or DM$^2$T. On the other portion of the fixed lines the FS is unstable to pairing, which may potentially be relevant for DMT or DM$^2$T. Finally, if the above scenario is true, it is worth investigating what underlying feature associated with the theory protects the fixed line even upon including higher order corrections.  
    
    \item \underline{Scenario B}: Upon including higher order corrections, the fixed lines of the two-patch theory collapse into fixed points. Broadly speaking, there are two distinct types of fixed points, depending on their tendency towards pairing (\ie stable vs. unstable). 
    
    The stable fixed points characterize new examples of quantum spin liquids, as discussed above in `scenario A'. While interesting in their own right, they are not relevant for describing continous DMT/DM$^2$T.
    
    On the other hand, the nature of the unstable fixed points are determined by what state the boson $B$ is in, as elaborated in Sec. \ref{subsec: insulators}. These states can potentially be relevant for describing continuous DMT/DM$^2$T from metals into various insulating states without any neutral FS, or, to other distinct metals. However, we emphasize that even if such fixed points are found, the criticality associated with the putative critical points need to be examined carefully to ascertain whether they remain continuous.
    
    \item \underline{Scenario C}: After higher order corrections are included, there is no fixed point at all. This suggests that the two-patch theory actually does not describe a critical system, and the system is gapped for reasons other than pairing. For example, it is possible that one of the gauge couplings flows to infinity, which suggests that this gauge field confines within the two-patch theory. In this scenario, the nature of the resulting state is determined by other microscopic details, which are beyond the purview of the present discussion.
    
\end{enumerate}

\acknowledgements

We thank Zhen Bi, Sung-Sik Lee, Max Metlitski, T. Senthil, Dam Thanh Son and Chong Wang for useful discussions. LZ is supported by the John Bardeen Postdoctoral Fellowship at Perimeter Institute. Research at Perimeter Institute is supported in part by the Government of Canada through the Department of Innovation, Science and Economic Development Canada and by the Province of Ontario through the Ministry of Colleges and Universities. DC is supported by startup funds at Cornell University.

{\it Note added:} While this manuscript was being prepared for submission, an independent work appeared on the arXiv \cite{SS20}, which also attempts to propose a related but distinct critical theory in a different setting to describe a transition between a FL and FL*.

\bibliography{DeconfinedMottTransition.bib}

\clearpage
\onecolumngrid
\appendix

\section{Details of the $U(2)$ parton construction} \label{app: U(2) parton}

In this appendix we provide additional details of the $U(2)$ parton construction given by Eq. \eqref{eq: parton}. Here we focus on the case of a single electronic orbital, and it is straightforward to generalize the discussion to a lattice of orbitals. Just as a reminder, we denote the electron operator by $c_a$, where $a=1,2$ stands for the two spin d.o.f. We will consider the following parton construction:
\beq \label{eq: U(2) parton for electron}
\left(
\begin{array}{c}
	c_1\\
	c_2
\end{array}
\right)
=
\left(
\begin{array}{cc}
	B_{11} & B_{12} \\
	B_{21} & B_{22}
\end{array}
\right)
\cdot
\left(
\begin{array}{c}
	f_1\\
	f_2
\end{array}
\right)
\eeq
where $f_{1,2}$ are complex fermion operators and $B$'s are canonical boson operators. Schematically, this equation reads $c=Bf$, where $c$ is a 2-by-1 matrix, $B$ is a 2-by-2 matrix and $f$ is a 2-by-1 matrix.

Apparently this parton construction has a $U(2)$ gauge redundancy, and the gauge transformation reads
\beq
\left(
\begin{array}{cc}
	B_{11} & B_{12} \\
	B_{21} & B_{22}
\end{array}
\right)
\rightarrow
\left(
\begin{array}{cc}
	B_{11} & B_{12} \\
	B_{21} & B_{22}
\end{array}
\right)\cdot U^\dag,
\quad
\left(
\begin{array}{c}
	f_1\\
	f_2
\end{array}
\right)
\rightarrow
U\cdot
\left(
\begin{array}{c}
	f_1\\
	f_2
\end{array}
\right)
\eeq
where $U$ is a $U(2)$ matrix. The $U(2)$ transformation has 4 generators: $Q$, the generator of the $U(1)$ part, and $T_{x,y,z}$, the generators of the $SU(2)$ part. The explicit expressions of these generators can be written in terms the $f$ and $b$ operators:
\beq
\begin{split}
	&Q=f_1^\dag f_1+f_2^\dag f_2-B_{11}^\dag B_{11}-B_{12}^\dag B_{12}-B_{21}^\dag B_{21}-B_{22}^\dag B_{22}=f^\dag f-\Tr(B^\dag B)\\
	&T_x=\frac{1}{2}\left(f_1^\dag f_2+f_2^\dag f_1-B_{11}^\dag B_{12}-B_{12}^\dag B_{11}-B_{21}^\dag B_{22}-B_{22}^\dag B_{21}\right)=\frac{1}{2}f^\dag\sigma_x f-\frac{1}{2}\Tr(B^\dag B\sigma_x)\\
	&T_y=-\frac{i}{2}\left(f_1^\dag f_2-f_2^\dag f_1+B_{11}^\dag B_{12}-B_{12}^\dag B_{11}+B_{21}^\dag B_{22}-B_{22}^\dag B_{21}\right)=\frac{1}{2}f^\dag \sigma_yf-\frac{1}{2}\Tr(B^\dag B\sigma_y)\\
	&T_z=\frac{1}{2}\left(f_1^\dag f_1-f_2^\dag f_2+B_{12}^\dag B_{12}+B_{22}^\dag B_{22}-B_{11}^\dag B_{11}-B_{21}^\dag B_{21}\right)=\frac{1}{2}f^\dag \sigma_z f-\frac{1}{2}\Tr(B^\dag B\sigma_z)
\end{split}
\eeq
with $\sigma$'s the standard Pauli matrices.

Demanding that the theory be in a gauge neutral sector, or equivalently, demanding that $Q=T_x=T_y=T_z=0$,  yields the gauge constraints:
\beq \label{eq: U(2) gauge constraints for electrons}
\begin{split}
	&B_{11}^\dag B_{11}+B_{21}^\dag B_{21}-f_1^\dag f_1=0\\
	&B_{12}^\dag B_{12}+B_{22}^\dag B_{22}-f_2^\dag f_2=0\\
	& B_{12}^\dag B_{11}+B_{22}^\dag B_{21}-f_1^\dag f_2=0\\
	& B_{11}^\dag B_{12}+B_{21}^\dag B_{22}-f_2^\dag f_1=0.
\end{split}
\eeq
It is straightforward to check that on a single orbital only 4 states satisfy all these constraints, and they are $|0\ra$, $|1\ra\equiv |B_{11}f_1\ra+|B_{12}f_2\ra$, $|2\ra\equiv|B_{21}f_1\ra+|B_{22}f_2\ra$ and $|d\ra\equiv|B_{11}B_{22}f_1f_2\ra-|B_{12}B_{21}f_1f_2\ra$, where $|0\ra$ is a state with no $B$ or $f$ particle, and $|B_{ij}f_k\ra\equiv B_{ij}^\dag f_k^\dag|0\ra$ for $i, j, k=1,2$, and $|B_{i_1i_2}B_{j_1j_2}f_{k}f_l\ra\equiv B^\dag_{i_1i_2}B^\dag_{j_1j_2}f_k^\dag f_l^\dag|0\ra$. Not surprisingly, if we identify $|0\ra$ as the empty site, then $|1\ra=c_1^\dag|0\ra$ has a single spin-up electron, $|2\ra=c_2^\dag|0\ra$ has a single spin-down electron, and $|d\ra=c_1^\dag c_2^\dag|0\ra$ is a doubly-occupied state that has both spin-up and spin-down electrons. Therefore, the parton construction Eq. \eqref{eq: U(2) parton for electron} supplemented with gauge constraints Eq. \eqref{eq: U(2) gauge constraints for electrons} is a faithful representation of all of the physical electronic states.

Furthermore, as noted in the main text, various global symmetries can be implemented as
\beq
\begin{split}
	&U(1): B\rightarrow e^{i\theta}B,\quad f\rightarrow f\\
	&SU(2):
	B
	\rightarrow VB,
	\quad
	f
	\rightarrow
	f\\
	& \mc{T}:
	B
	\rightarrow \epsilon B,
	\quad
	f
	\rightarrow
	f\\
\end{split}
\eeq
where $V$ is an $SU(2)$ matrix and $\epsilon$ is the rank-2 anti-symmetric tensor with $\epsilon_{12}=-\epsilon_{21}=1$. For later convenience, we write down the physical $U(1)$ charge,
\beq
\nu=B_{11}^\dag B_{11}+B_{12}^\dag B_{12}+B_{21}^\dag B_{21}+B_{22}^\dag B_{22}=\Tr(B^\dag B)
\eeq
from which we see that the density of $B$ is the same as the density of the physical electrons. The generators of the SU(2) spin rotational symmetry:
\beq
\begin{split}
	&S_x=\frac{1}{2}\left(B_{11}^\dag B_{21}+B_{21}^\dag B_{11}+B_{12}^\dag B_{22}+B_{22}^\dag B_{12}\right)=\frac{1}{2}\Tr(B^\dag\sigma_x B)\\
	&S_y=\frac{-i}{2}\left(B_{11}^\dag B_{21}-B_{21}^\dag B_{11}+B_{12}^\dag B_{22}-B_{22}^\dag B_{12}\right)=\frac{1}{2}\Tr(B^\dag\sigma_yB)\\
	&S_z=\frac{1}{2}\left(B_{11}^\dag B_{11}-B_{21}^\dag B_{21}+B_{12}^\dag B_{12}-B_{22}^\dag B_{22}\right)=\frac{1}{2}\Tr(B^\dag \sigma_zB).
\end{split}
\eeq

\section{Topological nature of some examples of insulating phases} \label{app: type-2}

In this appendix we apply the method in Ref. \cite{Zou2018} (see Appendix A therein) to derive the topological nature of a few examples of the insulating phases described in Sec. \ref{subsec: insulators}.

\subsection{Type-I insulator: short-range entangled insulator}

Let us start with the type-I state. Recall that in this state $B$ is in a completely trivial state, and $f$ is paired in the singlet $s$-wave channel of the $SU(2)$ gauge group. The low-energy topological quantum field theory of the system is given by
\beq
\mc{L}=-\frac{1}{\pi}\tilde adb
\eeq
where $\tilde a$ is the original dynamical $U(1)$ gauge field, and $b$ is a gauge field such that $\frac{1}{2\pi}db$ is the current of the singlet Cooper pairs of $f$. This topological action comes from pairing of the fermions in the singlet channel of the $SU(2)$ gauge field, $a$.

To understand the topological properties of such a theory, it is crucial to understand the possible charged excitations of these gauge fields. In our case, any excitation with an odd (even) charge under $\tilde a$ must carry a half-odd-integer (integer) spin under $a$, and vice versa. Also, due to the presence of local electrons, without loss of generality, in analyzing the topological properties of this theory, we can assume that the charges of these gauge fields are all bosonic. 

Now we note that because of the confinement of the $SU(2)$ gauge field, all excitations should be a singlet under $SU(2)$, which must carry even charge under $\tilde a$. More precisely, in terms of $\tilde a'=2\tilde a$, the charges of all the excitations take all integers. Therefore, the above topological theory is more appropriately written as
\beq
\mc{L}=-\frac{1}{2\pi}\tilde a'db
\eeq
where there is no restriction on the charges of the excitations under $\tilde a'$ and $b$. This theory is topologically trivial, \ie short-range entangled \cite{Wen2004Book}.

\subsection{Chiral spin liquid as a type-II insulator}

Next, we turn to the example of the type-II state described in Sec. \ref{subsec: insulators}. Recall that in this insulator $B$ is in a $U(2)$ symmetric bosonic integer quantum Hall (BIQH) state discussed in Ref. \cite{Senthil2012}, and $f$ are gapped out due to pairing in the singlet $s$-wave channel of the $SU(2)$ gauge field.

The low-energy topological quantum field theory of this state is given by
\beq \label{eq: TQFT}
\mc{L}=\frac{2}{4\pi}\tilde ad\tilde a-\frac{1}{4\pi}\Tr\left(ada-\frac{2i}{3}a^3\right)-\frac{1}{\pi}\tilde adb
\eeq
where, again, $\tilde a$ is the original dynamical $U(1)$ gauge field, $a$ is the dynamical $SU(2)$ gauge field, and $b$ is a gauge field such that $\frac{1}{2\pi}db$ is the current of the singlet Cooper pairs of $f$. The first two terms represent the the response of the BIQH of $B$ to the dynamical $U(2)$ gauge field \cite{Liu2012, Senthil2012, Seiberg2016, Zou2018, Ning2019}, and the last term is due to the pairing of fermions in the singlet channel of $a$.

It is convenient to split the TQFT Eq. \eqref{eq: TQFT} into two parts, \ie $\mc{L}=\mc{L}_1+\mc{L}_2$, where
\beq
\begin{split}
    &\mc{L}_1=-\frac{1}{4\pi}\Tr\left(ada-\frac{2i}{3}a^3\right)\\
    &\mc{L}_2=\frac{2}{4\pi}\tilde ad\tilde a-\frac{1}{\pi}\tilde adb=\frac{1}{4\pi}(\tilde a, b)\cdot K\cdot (d\tilde a, db)^T
\end{split}
\eeq
with the $K$-matrix:
\beq
K=
\left(
\begin{array}{cc}
2 & -2 \\
-2 & 0
\end{array}
\right)
\eeq

The theory of $\mc{L}_1$ is $SU(2)_1$, which has a single anyonic excitation that is a semion, which has topological spin $i$. This semion can be created by creating an excitation in the spinor representation of $a$. Excitations obtained by creating a linear representation of $ a$ are all local bosons.

Now we move to determine the topological properties of the excitations in $\mc{L}_2$, which turns out to be equivalent to a double-semion topological order. The excitations in $\mc{L}_2$ can be labeled by $(l_1, l_2)$, where $l_1$ and $l_2$ are the charges of this excitation under $\tilde a$ and $b$, respectively. Consider only $\mc{L}_2$, the topological spin of this excitation is given by
\beq
\theta^{(0)}_{(l_1, l_2)}=\exp\left(-i\pi (l_1, l_2)\cdot K^{-1}\cdot(l_1, l_2)^T\right)=\exp\left(\frac{i\pi}{2}(2l_1l_2+l_2^2)\right)
\eeq
Let us enumerate a few of these topological spins:
\beq
\theta_{(1, 0)}^{(0)}=1,
\quad
\theta_{(0, 1)}^{(0)}=i,
\quad
\theta_{(1,1)}^{(0)}=\theta_{(1,-1)}^{(0)}=-i
\eeq
In fact, $(1, 0)$, $(0, 1)$ and $(1, 1)$ are just the bosonic, semionic and anti-semionic bosons of the double-semion topological order, respectively.

Because the charges under $a$ and $\tilde a$ are not independent, these two sectors are not decoupled. In fact, because an excitation with odd (even) $l_1$ carries a half-odd-integer (integer) spin under $a$, the topological spin of this excitation should be further multipied by $i$ (1). So the true topological spins of these excitations are
\beq
\theta_{(1, 0)}=i,
\quad
\theta_{(0, 1)}=i,
\quad
\theta_{(1,1)}=\theta_{(1,-1)}=1
\eeq
It appears from the above topological spins that there are two semions in the theory, $(1, 0)$ and $(0, 1)$. However, these two excitations should actually be identified. To see it, consider the excitation $(1, -1)$, which is the bound state $(1, 0)$ and the anti-particle of $(0, 1)$. Its braiding statistics with any excitation $(l_1, l_2)$ is
\beq
\theta^{(1, -1)}_{(l_1, l_2)}=\exp\left(i\pi l_1\right)\cdot\exp\left(-2\pi i (1, -1)\cdot K^{-1}\cdot(l_1, l_2)^T\right)
\eeq
where the first and second factors come from the $\mc{L}_1$ and $\mc{L}_2$ sectors, respectively. It is straightforward to check that $\theta^{(1, -1)}_{(l_1, l_2)}=1$ for any $l_1$ and $l_2$. That is, $(1, -1)$ braids trivially with all excitations. So it should be a local excitation, which also means that $(1, 0)$ and $(0, 1)$ are actually in the same topological sector \cite{Kitaev2006}, and they represent a single semionic excitation in the system. One can further check that this semion is the only anyonic excitation in the theory Eq. \eqref{eq: TQFT}.

The above discussion has established that the bulk topological order of the theory Eq. \eqref{eq: TQFT} is identical to a chiral spin liquid (also known as a Laughlin-$1/2$ state) \cite{Wen2004Book}. Now we examine the edge modes of this theory, which can differ by chiral Majorana modes for a given bulk topological order. Integrating out $a$, $\tilde a$ and $b$ generates a gravitational Chern-Simons term $\gcs$, which means that the system has a single chiral edge mode with chiral central charge $c_-=1$. Therefore, this topological order is indeed a chiral spin liquid state, with no additional chiral Majorana edge modes which will shift the gravitational Chern-Simons term by a multiple of $\gcs/2$.

\subsection{$Z_2$ topological order as a type-III insulator}

Now we give an example of a type-III state. Consider the case where $B$ itself is in the $Z_2$ topological order (with no additional nontrivial Chern-Simons response to the dynamical $U(2)$ gauge field), and $f$ is paired in the singlet $s$-wave channel of the $SU(2)$ gauge group. Furthermore, let us assume that the $Z_2$ topological order is completely neutral under $U(2)$ (\ie corresponding to the trivial fractionalization pattern of $U(2)$). The low-energy topological quantum field theory of the system can be written as
\beq
\mc{L}=-\frac{1}{\pi}\tilde adb+\frac{1}{\pi}a_1da_2
\eeq
As in the case of the type-I state, only even charge of $\tilde a$ is allowed, while there is no restriction on the charges of other gauge fields. Therefore, this topological theory can be more appropriately written as
\beq
\mc{L}=-\frac{1}{2\pi}\tilde a'db+\frac{1}{\pi}a_1da_2
\eeq
with $\tilde a'=2\tilde a$. This is still a $Z_2$ topological order.

\subsection{An example of a type-IV insulator}

Finally, we give a relatively simple example of a type-IV insulator. We simply stack the type-II and type-III states of $B$ described above. Similar arguments as above indicate that the system will be in a topologically ordered state that can be viewed as a chiral spin liquid stacked on top of a $Z_2$ topological order.

\section{Framework for Wilsonian Renormalization Group} \label{app: Wilsonian RG}

In this appendix, we generalize the methods developed in Refs. \cite{Metlitski2010, Mross2010} to describe the framework we use to carry out a Wilsonian RG analysis to the two-patch theory described by Eq. \eqref{eq: 2-U(1) patch}. Our derivation of the framework will also help clarify some of the conceptual points that arise in these calculations.

Suppose the theory is defined originally with all of the coupling constants at a  cutoff scale, $\Lambda$, by which we mean that all modes with $|k_y|<\Lambda$ are included in the theory, while other modes are not included. Suppose that after integrating out the fast modes with $|k_y|\in(\Lambda/\sqrt{b}, \Lambda)$, the effective action becomes (with the generated irrelevant perturbations discarded):
\beq \label{eq: effective action one-shell integrated}
\begin{split}
S_\eff=
&\int^{\frac{\sqrt{b}}{\Lambda}}d\tau dxdy
\Bigg[
\sum_{\alpha=1,2; p=\pm}\left(\left(1+\delta_f\right)\psi_{\alpha p}^\dag(-ip\partial_x-\partial_y^2)\psi_{\alpha p}+\left(\eta+\delta_\eta\right)\psi_{\alpha p}^\dag\partial_\tau \psi_{\alpha p}\right)\\
&\qquad\qquad\qquad\qquad
-\left(1+\delta_c\right)\sum_{p=\pm}\lambda_p a_c\left(\psi_{1p}^\dag \psi_{1p}+\psi_{2p}^\dag \psi_{2p}\right)-\left(1+\delta_s\right)\sum_{p=\pm}\lambda_p a_s\left(\psi_{1p}^\dag \psi_{1p}-\psi_{2p}^\dag \psi_{2p}\right)
\Bigg]\\
&+\int^{\frac{\Lambda}{\sqrt{b}}}\frac{d\omega dk_x dk_y}{(2\pi)^3}\frac{N|k_y|^{1+\epsilon}}{2}\left(\frac{1}{e_c^2+\delta_{e_c^2}}|a_c|^2+\frac{1}{e_s^2+\delta_{e_s^2}}|a_s|^2\right),
\end{split}
\eeq
where the superscript $\sqrt{b}/\Lambda$ or $\Lambda/\sqrt{b}$ indicates that only modes with $k_y<\Lambda/\sqrt{b}$ are included in the theory, and $\delta_f$, $\delta_{\eta}$, $\delta_c$, $\delta_{s}$, $\delta_{e_c^2}$ and $\delta_{e_s^2}$ arise as a result of integrating out the fast modes. Clearly, the $\delta$'s depend on $b$ and they approach zero when $b\rightarrow 1$. Furthermore, all these $\delta$'s are expected to have no explicit dependence on $\Lambda.$

In order to obtain the beta functions of various couplings, we need to rescale the frequency and momenta so that the cutoff is $\Lambda$ again, which requires us to determine the relative scaling between imaginary time and real space, specified by a ``dynamical exponent", $z_\Lambda$. The value of $z_\Lambda$ can be taken to be arbitrary for now, and this value determines the RG scheme we use, as will be clear soon. For now, we leave the value of $z_\Lambda$ unspecified until we decide on the RG scheme. We stress that this $z_\Lambda$ should not be confused with the physical dynamical exponent, which is obtained from the scaling structure of the correlation functions.
 
We define $\tau'=\tau b^{-z_\Lambda}$, $x'=xb^{-1}$ and $y'=yb^{-1/2}$, and we rewrite the effective action given by Eq. \eqref{eq: effective action one-shell integrated} as
\beq \label{eq: effective action after the whole package}
\begin{split}
S_\eff=
&\int^{\frac{1}{\Lambda}}d\tau'dx'dy'
\left[
\sum_{\alpha=1,2; p=\pm}\psi'^\dag_{\alpha p}(\eta'\partial_{\tau'}-ip\partial_{x'}-\partial_{y'}^2)\psi'_{\alpha p}-\sum_{p=\pm}\lambda_p\left((a_c'+a_s')\psi'^{\dag}_{1p} \psi'_{1p}+(a_c'-a_s')\psi'^\dag_{2p} \psi'_{2p}\right)
\right]\\
&
+\int^{\Lambda}\frac{d\omega' dk_x' dk_y'}{(2\pi)^3}\left(\frac{N}{2e'^2_c}|k'_y|^{1+\epsilon}|a'_c|^2+\frac{N}{2e'^2_s}|k'_y|^{1+\epsilon}|a'_s|^2\right).
\end{split}
\eeq
Comparing this effective action and Eq. \eqref{eq: effective action one-shell integrated} yields
\beq \label{eq: renormalized fields and couplings after the whole package}
\begin{split}
&\psi'_{\alpha p}=\sqrt{1+\delta_f}\cdot b^{\frac{z_\Lambda}{2}+\frac{1}{4}}\cdot \psi_{\alpha p},
\qquad
\eta'=\frac{\eta+\delta_\eta}{1+\delta_f}\cdot b^{1-z_\Lambda},\\
&a'_c=\frac{1+\delta_c}{1+\delta_f}\cdot b\cdot a_c,
\qquad
e'^2_c=(e_c^2+\delta_{e_c^2})\left(\frac{1+\delta_c}{1+\delta_f}\right)^2b^{-z_\Lambda+1+\frac{\epsilon}{2}},
\\
&a'_s=\frac{1+\delta_s}{1+\delta_f}\cdot b\cdot a_s,
\qquad
e'^2_s=(e_s^2+\delta_{e_s^2})\left(\frac{1+\delta_s}{1+\delta_f}\right)^2b^{-z_\Lambda+1+\frac{\epsilon}{2}}.
\end{split}
\eeq

Therefore, the beta functions for $\eta$, $e_c^2$ and $e_s^2$ are
\beq \label{eq: beta functions}
\begin{split}
&\beta(\eta)=\lim_{b\rightarrow 1}\frac{\eta'-\eta}{b-1}=(1-z_\Lambda)\eta+\lim_{b\rightarrow 1}\frac{\delta_\eta-\eta\delta_f}{b-1},\\
&\beta(e_c^2)=\lim_{b\rightarrow 1}\frac{e'^2_c-e_c^2}{b-1}=\left(1+\frac{\epsilon}{2}-z_\Lambda\right)e_c^2+\lim_{b\rightarrow 1}\frac{\delta_{e_c^2}}{b-1}+2e_c^2\lim_{b\rightarrow 1}\frac{\delta_c-\delta_f}{b-1},\\
&\beta(e_s^2)=\lim_{b\rightarrow 1}\frac{e'^2_s-e_s^2}{b-1}=\left(1+\frac{\epsilon}{2}-z_\Lambda\right)e_s^2+\lim_{b\rightarrow 1}\frac{\delta_{e_s^2}}{b-1}+2e_s^2\lim_{b\rightarrow 1}\frac{\delta_s-\delta_f}{b-1}.
\end{split}
\eeq
From the above expressions, we see that the beta functions depend on the choice of $z_\Lambda$, which is determined by the RG scheme. One choice is to take $z_\Lambda=1$ and another choice is to take the value of $z_\Lambda$ so that $\eta$ does not flow, \ie $\beta(\eta)=0$. Notice for all choices, $e_c^2/\eta$ and $e_s^2/\eta$ have a $z_\Lambda$-independent beta functions:
\beq
\begin{split}
	&\beta\left(\frac{e_c^2}{\eta}\right)=\frac{\epsilon}{2}\cdot\frac{e_c^2}{\eta}+\lim_{b\rightarrow 1}\frac{1}{b-1}\cdot\left[\frac{\delta_{e_c^2}+2e_c^2(\delta_c-\delta_f)}{\eta}-\frac{e_c^2(\delta_\eta-\eta\delta_f)}{\eta^2}\right]\\
	&\beta\left(\frac{e_s^2}{\eta}\right)=\frac{\epsilon}{2}\cdot\frac{e_s^2}{\eta}+\lim_{b\rightarrow 1}\frac{1}{b-1}\cdot\left[\frac{\delta_{e_s^2}+2e_s^2(\delta_c-\delta_f)}{\eta}-\frac{e_s^2(\delta_\eta-\eta\delta_f)}{\eta^2}\right].
\end{split}
\eeq
Notice $\delta_{e_c^2}=\delta_{e_s^2}$=0 if $\epsilon<1$, due to the non-analytic nature of the kinetic term of the gauge fields.

Often it is simpler to evaluate the $\delta$'s first and obtain the associated beta functions, as above. However, this approach has the disadvantage that the properties of the correlation functions of the theory are obscure. Therefore, we apply a different method which allows us to read off the beta functions directly from the correlation functions. Within this approach, the structure of the correlation functions is more transparent.

To this end, we will first derive a Callan-Symanzik type equation. For this purpose, let us first define
\beq \label{eq: renormalization factors}
\begin{split}
&\tilde \psi_{\alpha}\equiv Z_f^{\frac{1}{2}}\psi_{\alpha}\equiv\sqrt{1+\delta_f}\psi_{\alpha p}\\
&\tilde\eta\equiv \frac{\eta+\delta_{\eta}}{1+\delta_f}\\
&\tilde a_c\equiv Z_c^{\frac{1}{2}}\cdot a_c\equiv\frac{1+\delta_c}{1+\delta_f}\cdot a_c\\
&\tilde a_s\equiv Z_s^{\frac{1}{2}}\cdot a_s\equiv\frac{1+\delta_s}{1+\delta_f}\cdot a_s\\
&\tilde e_c^2\equiv Z_{e_c^2}\cdot e_c^2\equiv\left(\frac{1+\delta_c}{1+\delta_f}\right)^2\cdot\left(e_c^2+\delta_{e_c^2}\right)\\
&\tilde e_s^2\equiv Z_{e_s^2}\cdot e_s^2\equiv\left(\frac{1+\delta_s}{1+\delta_f}\right)^2\cdot\left(e_s^2+\delta_{e_s^2}\right),
\end{split}
\eeq
so that the effective action, Eq. \eqref{eq: effective action one-shell integrated}, in terms of these fields and quantities with tildes, reads
\beq \label{eq: effective action after a partial rescaling}
\begin{split}
	S_\eff=
	&\int^{1/\tilde\Lambda}d\tau dxdy
	\Bigg[
	\sum_{\alpha=1,2; p=\pm}\tilde \psi^\dag_{\alpha p}(\tilde\eta\partial_{\tau}-ip\partial_{x}-\partial_{y}^2)\tilde \psi_{\alpha p}-\sum_{p=\pm}\lambda_p\left((\tilde a_c+\tilde a_s)\tilde \psi^{\dag}_{1p} \tilde \psi_{1p}+(\tilde a_c-\tilde a_s)\tilde \psi^\dag_{2p} \tilde \psi_{2p}\right)
	\Bigg]\\
	&
	+\int^{\tilde\Lambda}\frac{d\omega dk_x dk_y}{(2\pi)^3}\left(\frac{N}{2\tilde e^2_c}|k_y|^{1+\epsilon}|\tilde a_c|^2+\frac{N}{2\tilde e^2_s}|k_y|^{1+\epsilon}|\tilde a_s|^2\right),
\end{split}
\eeq
with $\tilde\Lambda\equiv\Lambda/\sqrt{b}$. Notice that this form of the effective action differs from Eq. \eqref{eq: effective action after the whole package} in that the space-time is not rescaled to match the new cutoff with the original one, while the forms of the Lagrangian in Eqs. \eqref{eq: 2-U(1) patch}, \eqref{eq: effective action after the whole package} and \eqref{eq: effective action after a partial rescaling} are the same.

Next, consider a physical correlation function, $\Gamma^{n_f, n_c, n_s}$, made of $n_f$ fermion fields, $n_c$ $a_c$-fields and $n_s$ $a_s$-fields. Denote the corresponding correlation function calculated in terms of the fields $\tilde\psi$, $\tilde a_c$ and $\tilde a_s$ by $\tilde\Gamma^{n_f, n_c, n_s}$, which is related to $\Gamma^{n_f, n_c, n_s}$ via
\beq \label{eq: bare and renormalized correlators}
\Gamma^{n_f, n_c, n_s}=Z_f^{-\frac{n_f}{2}}Z_c^{-\frac{n_c}{2}}Z_s^{-\frac{n_s}{2}}\cdot\tilde\Gamma^{n_f, n_c, n_s}.
\eeq
In general, $\Gamma^{n_f, n_c, n_s}$ is a function of $\tilde\Lambda$, $\tilde e_c^2$, $\tilde e_s^2$, $\tilde\eta$, $Z_f$, $Z_c$ and $Z_s$. Similarly, $\tilde\Gamma^{n_f, n_c, n_s}$ is in general a function of $\tilde\Lambda$, $\tilde e_c^2$, $\tilde e_s^2$ and $\tilde\eta$. Notice, however, there is {\em no} explicit dependence of $\tilde\Gamma^{n_f, n_c, n_s}$ on $Z_f$, $Z_c$ or $Z_s$.

When the cutoff varies (\eg from $\Lambda$ to $\tilde\Lambda$), in order to keep the physical correlation function, $\Gamma^{n_f, n_c, n_s}$, invariant, the values of $\tilde e_c^2$, $\tilde e_s^2$, $\tilde\eta$, $Z_f$, $Z_c$ and $Z_s$ also need to be adjusted accordingly. The invariance of $\Gamma^{n_f, n_c, n_s}$ yields a constraint equation between the changes of $\Lambda$, $\tilde e_c^2$, etc. This equation is the Callan-Symanzik equation that will be derived below.

For notational convenience, let us define
\beq \label{eq: renormalization factors after partial rescaling}
\begin{split}
	&b_c=\frac{\tilde\Lambda}{\tilde e_c^2}\frac{d\tilde e_c^2}{d\tilde\Lambda}=\lim_{b\rightarrow 1}\frac{\Lambda}{\tilde e_c^2}\frac{\tilde e_c^2-e_c^2}{\frac{\Lambda}{\sqrt{b}}-\Lambda}=-\frac{2}{e_c^2}\lim_{b\rightarrow 1}\frac{\delta_{e_c^2}+2e_c^2\left(\delta_c-\delta_f\right)}{b-1}\\
	&b_s=\frac{\tilde\Lambda}{\tilde e_s^2}\frac{d\tilde e_s^2}{d\tilde\Lambda}=\lim_{b\rightarrow 1}\frac{\Lambda}{\tilde e_s^2}\frac{\tilde e_s^2-e_s^2}{\frac{\Lambda}{\sqrt{b}}-\Lambda}=-\frac{2}{e_s^2}\lim_{b\rightarrow 1}\frac{\delta_{e_s^2}+2e_s^2\left(\delta_s-\delta_f\right)}{b-1}\\
	&b_\eta=\tilde\Lambda\frac{d\log\tilde\eta}{d\tilde\Lambda}=\lim_{b\rightarrow 1}\frac{\Lambda}{\eta}\frac{\frac{\eta+\delta_{\eta}}{1+\delta_f}-1}{\frac{\Lambda}{\sqrt{b}}-\Lambda}=-\frac{2}{\eta}\cdot\lim_{b\rightarrow 1}\frac{\delta_{\eta}-\eta\delta_f}{b-1}\\
	&\gamma_f=\frac{\tilde\Lambda}{Z_f}\frac{dZ_f}{d\tilde\Lambda}=\lim_{b\rightarrow 1}\frac{\Lambda}{Z_f}\frac{\delta_f}{\frac{\Lambda}{\sqrt{b}}-\Lambda}=-2\cdot\lim_{b\rightarrow 1}\frac{\delta_f}{b-1}\\
	&\gamma_c=\frac{\tilde\Lambda}{Z_c}\frac{d Z_c}{d\tilde\Lambda}=2\cdot\lim_{b\rightarrow 1}\frac{\Lambda}{Z_c}\frac{\delta_c-\delta_f}{\frac{\Lambda}{\sqrt{b}}-\Lambda}=-4\cdot\lim_{b\rightarrow 1}\frac{\delta_c-\delta_f}{b-1}\\
	&\gamma_s=\frac{\tilde\Lambda}{Z_s}\frac{d Z_s}{d\tilde\Lambda}=2\cdot\lim_{b\rightarrow 1}\frac{\Lambda}{Z_s}\frac{\delta_s-\delta_f}{\frac{\Lambda}{\sqrt{b}}-\Lambda}=-4\cdot\lim_{b\rightarrow 1}\frac{\delta_s-\delta_f}{b-1}.
\end{split}
\eeq
Notice that using some of the above quantities, the various beta functions introduced before can be written as
\beq \label{eq: beta functions simpler-looking}
\begin{split}
	&\beta(\eta)=\left(1-z_\Lambda-\frac{b_\eta}{2}\right)\cdot\eta\\
	&\beta(e_c^2)=\left(1+\frac{\epsilon}{2}-z_\Lambda-\frac{b_c}{2}\right)e_c^2\\
	&\beta(e_s^2)=\left(1+\frac{\epsilon}{2}-z_\Lambda-\frac{b_s}{2}\right)e_c^2\\
	&\beta(e_c^2/\eta)=\frac{\epsilon+b_\eta-b_c}{2}\cdot\frac{e_c^2}{\eta}\\
	&\beta(e_s^2/\eta)=\frac{\epsilon+b_\eta-b_s}{2}\cdot\frac{e_s^2}{\eta}.
\end{split}
\eeq
From these equations we see immediately that in order to obtain the beta functions, all we need are $b_c$, $b_s$ and $b_\eta$. 

The invariance of $\Gamma^{n_f, n_c, n_s}$ implies that
$\tilde\Lambda\frac{d}{d\tilde\Lambda}\Gamma^{n_f, n_c, n_s}=0$. More explicitly,
\beq \label{eq: almost Callan-Symanzik equation for 2-U(1)}
\begin{split}
	&\left(\tilde\Lambda\frac{\partial}{\partial\tilde\Lambda}+b_c\tilde e_c^2\frac{\partial}{\partial\tilde e_c^2}+b_s\tilde e_s^2\frac{\partial}{\partial\tilde e_s^2}+b_\eta\tilde\eta\frac{\partial}{\partial\tilde\eta}+\gamma_f\cdot Z_f\frac{\partial}{\partial Z_f}+\gamma_c\cdot Z_c\frac{\partial}{\partial Z_c}+\gamma_s\cdot Z_s\frac{\partial}{\partial Z_s}\right)\cdot\\
	&\qquad\qquad\qquad\qquad\qquad\qquad\qquad
	\Gamma^{n_f, n_c, n_s}\left(\{p_y\}, \{p_x\}, \{\omega\}, \tilde e_c^2, \tilde e_s^2, \tilde\eta, Z_f, Z_c, Z_s, \tilde\Lambda\right)=0.
\end{split}
\eeq
Now plugging Eq. \eqref{eq: bare and renormalized correlators} into the above equation and noting that $\tilde\Gamma^{n_f, n_c, n_s}$ has no explicit dependence on $Z_f$, $Z_c$ and $Z_s$, we finally arrive at the Callan-Symanzik equation:
\beq \label{eq: Callan-Symanzik equation for 2-U(1)}
\begin{split}
	&\left(\tilde\Lambda\frac{\partial}{\partial\tilde\Lambda}+b_c\tilde e_c^2\frac{\partial}{\partial \tilde e_c^2}+b_s\tilde e_s^2\frac{\partial}{\partial\tilde e_s^2}+b_\eta\tilde\eta\frac{\partial}{\partial\tilde\eta}-\frac{n_f}{2}\gamma_f-\frac{n_c}{2}\gamma_c-\frac{n_s}{2}\gamma_s\right)\cdot\\
	&\qquad\qquad\qquad\qquad\qquad\qquad\qquad
	\tilde\Gamma^{n_f, n_c, n_s}\left(\{p_y\}, \{p_x\}, \{\omega\}, \tilde e_c^2, \tilde e_s^2, \tilde\eta, \tilde\Lambda\right)=0,
\end{split}
\eeq

This Callan-Symanzik equation can be a lot more illuminating if the correlation function $\tilde\Gamma$ is expressed as a product of a dimensionful part and a dimensionless part. Using dimensional analysis, we find that $\tilde\Gamma$ can be written as
\beq \label{eq: decomposed correlator}
\tilde\Gamma^{n_f, n_c, n_s}=\tilde\Lambda^{\delta}\cdot g\left(\left\{\frac{p_y}{\tilde\Lambda}\right\}, \left\{\frac{p_x}{\tilde\Lambda^2}\right\}, \left\{\frac{\omega\tilde e_c^2}{\tilde\Lambda^{2+\epsilon}}\right\}, \left\{\frac{\omega\tilde e_s^2}{\tilde\Lambda^{2+\epsilon}}\right\}, \left\{\frac{\tilde\eta\omega}{\tilde\Lambda^2}\right\}, \left\{\frac{\omega}{\tilde\Lambda^z}\right\}\right)
\eeq
with 
\beq
\begin{split}
	\delta=z+3-\frac{z+5}{2}\cdot n_f-(z+1)\cdot(n_c+n_s).
\end{split}
\eeq
Notice that the value of $z$ can only be determined after obtaining the explicit expression of the correlation function and writing it in the above form, Eq. \eqref{eq: decomposed correlator}. Furthermore, {\it a priori}, the value of $z$ can be different for different correlation functions.

Plugging Eq. \eqref{eq: decomposed correlator} into Eq. \eqref{eq: Callan-Symanzik equation for 2-U(1)} yields another scaling form of the Callan-Symanzik equation
\beq \label{eq: Callan-Symanzik dimensionless}
\begin{split}
	&\left\{\frac{p_y}{\tilde\Lambda}\right\}g_1+2\cdot\left\{\frac{p_x}{\tilde\Lambda^2}\right\}g_2\\
	&+\left(2+\epsilon-b_c\right)\left\{\frac{\omega\tilde e_c^2}{\tilde\Lambda^{2+\epsilon}}\right\}g_3+\left(2+\epsilon-b_s\right)\left\{\frac{\omega\tilde e_s^2}{\tilde\Lambda^{2+\epsilon}}\right\}g_4+\left(2-b_\eta\right)\left\{\frac{\tilde\eta\omega}{\tilde\Lambda^2}\right\}g_5+z\left\{\frac{\omega}{\tilde\Lambda^z}\right\}g_6\\
	&
	=\left(\delta-\frac{n_f\gamma_f+n_c\gamma_c+n_s\gamma_s}{2}\right)g,
\end{split}
\eeq
with $g_i$ the partial derivative of $g$ with respect to its $i$-th argument, \ie $g_i\equiv\frac{\partial g(x_1, x_2, x_3, x_4, x_5, x_6)}{\partial x_i}$. From here we also see that $\gamma_f$, $\gamma_c$ and $\gamma_s$ are the anomalous dimensions of $f$, $a_c$ and $a_s$, respectively. Because of the Ward identities, $\gamma_c=\gamma_s=0$ for this $U_c(1)\times U_s(1)$ problem \cite{Metlitski2014}. Combining this, Eq. \eqref{eq: beta functions} and that $\delta_{e_c^2}=\delta_{e_s^2}=0$ when $\epsilon<1$ yields
\beq
\begin{split}
    &\beta(e_c^2)=\left(1+\frac{\epsilon}{2}-z_\Lambda\right)e_c^2\\
    &\beta(e_s^2)=\left(1+\frac{\epsilon}{2}-z_\Lambda\right)e_s^2
\end{split}
\eeq
which implies, when $\epsilon<1$, $\beta\left(e_c^2/e_s^2\right)=0$ to all orders, \ie $e_c^2/e_s^2$ is exactly RG invariant. On the other hand, although in the main text we argue that the RG framework developed here can be mostly applied to the $U(2)$ gauge theory as well, we stress that for the $U(2)$ gauge theory one cannot directly conclude that the ratio of the $U(1)$ and $SU(2)$ gauge couplings is exactly RG invariant when $\epsilon<1$, due to a possible nonzero anomalous dimension of the $SU(2)$ gauge field \cite{Peskin1995}.

We remark that $\tilde\Gamma^{n_f, n_c, n_s}$ is calculated from the theory defined with cutoff $\tilde\Lambda$ and coupling constants $\tilde e_c$, $\tilde e_s$ and $\tilde\eta$. By calculating it and requiring that it satisfy Eq. \eqref{eq: Callan-Symanzik equation for 2-U(1)} or Eq. \eqref{eq: Callan-Symanzik dimensionless}, we can obtain the values of $b_c$, $b_s$, $b_\eta$ and $\gamma_{f,c,s}$. Then we can substitute these quantities into Eq. \eqref{eq: beta functions simpler-looking} and get the corresponding beta functions. Notice, however, in order to obtain the physical correlation function, $\Gamma^{n_f, n_c, n_s}$, one still needs to substitute $\gamma_{f,c,s}$ into Eq. \eqref{eq: renormalization factors after partial rescaling} to solve for $Z_{f,c,s}$, and use Eq. \eqref{eq: bare and renormalized correlators}.

\section{Correlation functions and beta functions from Wilsonian RG} \label{app: fermion propagator}

Based on the framework for Wilsonian RG presented in Appendix \ref{app: Wilsonian RG}, here we give examples of some specific correlation functions, $\tilde\Gamma^{n_f, n_c, n_s}$, and obtain the beta functions to $\mc{O}(\epsilon)\sim\mc{O}(1/N)$.

\begin{figure}[h]
\begin{center}
\includegraphics[scale=0.08]{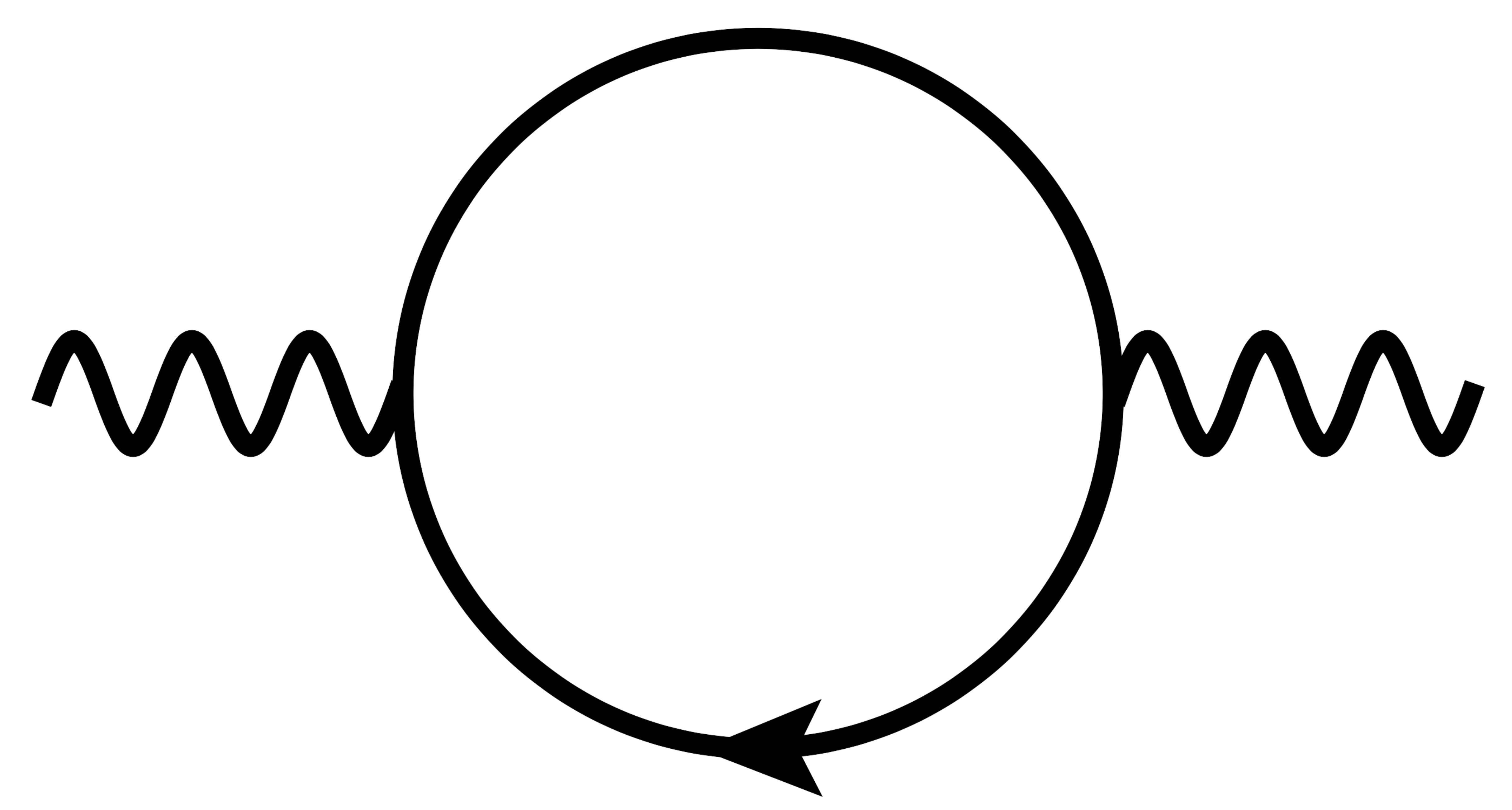}
\end{center}
\caption{The self-energy diagram of the gauge field. The external wavy line can represent any one of the gauge fields ($a_c$ and $a_s$ in the $U(1)\times U(1)$ gauge theory, or, $a$ and $\tilde a$ of the $U(2)$ gauge theory) described in this paper.}
\label{BSE}
\end{figure}

The propagator for $a_c$ at this order is \cite{Sachdev_book2011}
\beq
D_c(\omega, k_y)=\frac{1/N}{\gamma\frac{|\omega|}{|k_y|}+\frac{|k_y|^{1+\epsilon}}{e_c^2}},
\eeq
with $\gamma=1/(4\pi)$. This is obtained by adding the bare propagator and the self-energy shown in Fig. \ref{BSE}. Requiring that this propagator satisfy Eq. \eqref{eq: Callan-Symanzik equation for 2-U(1)} yields $b_c=0$. Similar considerations for the propagator of $a_s$ at this leading order will give $b_s=0$.

\begin{figure}[h]
\begin{center}
\includegraphics[scale=0.08]{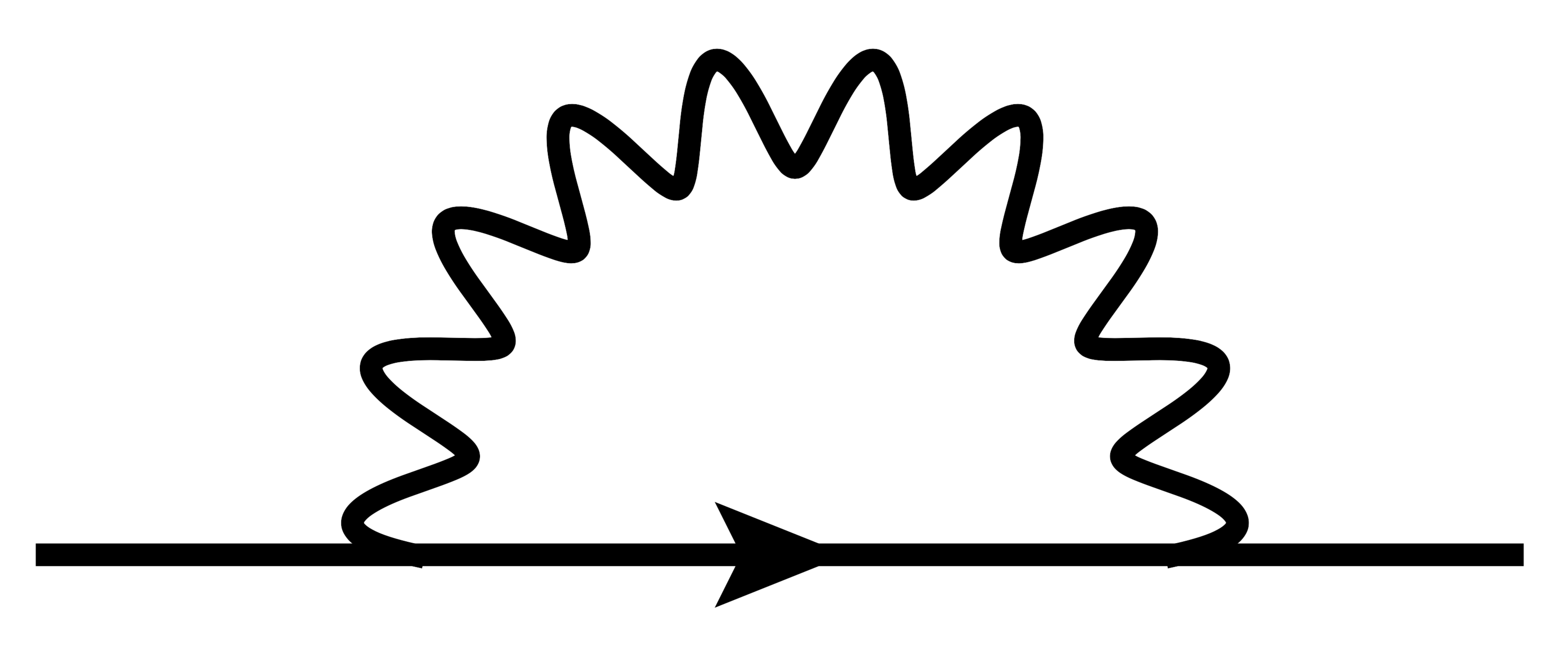}
\end{center}
\caption{The fermion self-energy diagram due to one of the gauge fields. In the $U(1)\times U(1)$ gauge theory, this diagram represents contribution from either $a_c$ or $a_s$. In the $U(2)$ gauge theory, it represents contribution from either $a$ or $\tilde a$. The total self-energy includes contributions from both gauge fields.}
\label{FSE}
\end{figure}

Next consider the self-energy of the fermion (see Fig. \ref{FSE} for the Feynman diagram and the end of this appendix for computational details):
\beq \label{eq: fermion propagator}
\begin{split}
	\Sigma_p(\omega, k_x, k_y)
	&=\int_{\Omega, q_x, q_y}(D_c(\Omega, q_y)+D_s(\Omega, q_y))G_{p}^{(0)}(\omega-\Omega, k_x-q_x, k_y-q_y)\\
	&=\frac{i}{4\pi N}\left(e_c^2(\gamma e_c^2)^{-\frac{\epsilon}{2+\epsilon}}+e_s^2(\gamma e_s^2)^{-\frac{\epsilon}{2+\epsilon}}\right)\sin^{-1}\left(\frac{2\pi}{2+\epsilon}\right)|\omega|^{\frac{2}{2+\epsilon}}\sgn(\omega)-\frac{i(e_c^2+e_s^2)}{2\pi^2N}\cdot\frac{\Lambda^{-\epsilon}\omega}{\epsilon}.
\end{split}
\eeq
Substituting this self-energy into the fermion propagator and using the Callan-Symanzik equation, Eq. \eqref{eq: Callan-Symanzik equation for 2-U(1)}, yield $b_\eta=-\frac{e_c^2+e_s^2}{2\pi^2N\eta\Lambda^\epsilon}+\mc{O}(\epsilon/N)$ and $\gamma_f=0+\mc{O}(\epsilon/N)$.

In passing, we note that the vertex correction to the fermion-gauge coupling vanishes at this order \cite{Mross2010}, as expected from the momentum-independence of fermion self-energy in Eq. \eqref{eq: fermion propagator} and the Ward identity of this $U(1)\times U(1)$ gauge theory \cite{Metlitski2010}. 

In summary, at $\mc{O}(\epsilon)\sim\mc{O}(1/N)$, the relevant RG factors are
\beq \label{eq: RG factors Wilsonian 2-U(1)}
\begin{split}
	&b_\eta=-\frac{e_c^2+e_s^2}{2\pi^2N\eta\Lambda^\epsilon},
	\quad
	b_c=b_s=0,\\
	&\gamma_f=\gamma_c=\gamma_s=0.
\end{split}
\eeq
Combining these results with Eq. \eqref{eq: beta functions simpler-looking} and choosing $z_\Lambda=1$ as in Ref. \cite{Metlitski2014} yield beta functions
\beq \label{eq: beta functions skip point}
\begin{split}
    &\beta(\eta)=\frac{e_c^2+e_s^2}{4\pi^2\eta\Lambda^\epsilon}\cdot\eta,\\
	&\beta(e_c^2)=\left(\frac{\epsilon}{2}-\frac{e_c^2+e_s^2}{4\pi^2 N\eta\Lambda^\epsilon}\right)\cdot e_c^2,\\
	&\beta(e_s^2)=\left(\frac{\epsilon}{2}-\frac{e_c^2+e_s^2}{4\pi^2 N\eta\Lambda^\epsilon}\right)\cdot e_s^2.
\end{split}
\eeq
Defining the dimensionless gauge coupling constants, $\alpha_c=\frac{e_c^2}{4\pi^2\eta\Lambda^\epsilon},~
\alpha_s=\frac{e_s^2}{4\pi^2\eta\Lambda^\epsilon}$, their corresponding beta functions are
\beq
\begin{split}
&\beta(\alpha_c)=\left(\frac{\epsilon}{2}-\frac{\alpha_c+\alpha_s}{N}\right)\alpha_c,\\
&\beta(\alpha_s)=\left(\frac{\epsilon}{2}-\frac{\alpha_c+\alpha_s}{N}\right)\alpha_s.
\end{split}
\eeq

Analogous calculations can be carried out for the $U(2)$ problem in the ``quasi-Abelianized" limit, where the propagator of $\tilde a$ is
\beq
\tilde D(\omega, k_y)=\frac{1/N}{\gamma\frac{|\omega|}{|k_y|}+\frac{|k_y|^{1+\epsilon}}{e^2}},
\eeq
and the propagator of $a^\alpha$ is
\beq
D^{\alpha\beta}(\omega, k_y)=\frac{2/N}{\gamma\frac{|\omega|}{|k_y|}+\frac{|k_y|^{1+\epsilon}}{g^2}}\cdot\delta^{\alpha\beta}.
\eeq
The self-energy of the fermions is $\Sigma_p^{\alpha\beta}(\omega, k_x, k_y)=\Sigma_p(\omega)\delta^{\alpha\beta}$, with
\beq \label{eq: fermion propagator U(2)}
\Sigma_p(\omega)=\frac{i}{4\pi N}\left(e^2(\gamma e^2)^{-\frac{\epsilon}{2+\epsilon}}+\frac{3}{2}g^2(2\gamma g^2)^{-\frac{\epsilon}{2+\epsilon}}\right)\sin^{-1}\left(\frac{2\pi}{2+\epsilon}\right)|\omega|^{\frac{2}{2+\epsilon}}\sgn(\omega)-\frac{i\left(e^2+\frac{3}{2}g^2\right)}{2\pi^2N}\cdot\frac{\Lambda^{-\epsilon}\omega}{\epsilon}.
\eeq
Moreover, the vertex correction to the fermion-gauge coupling vanishes at this order \cite{Mross2010}. Therefore at this order, the $U(2)$ analog of the quantities in Eq. \eqref{eq: RG factors Wilsonian 2-U(1)} are
\beq
\begin{split}
	&b_\eta=-\frac{e_c^2+\frac{3}{2}g^2}{2\pi^2N\eta\Lambda^\epsilon},
	\quad
	\tilde b=b=0,\\
	&\gamma_f=\tilde\gamma=\gamma=0.
\end{split}
\eeq
Similar to the $U_c(1)\times U_s(1)$ case, if we take $z_\Lambda=1$, we obtain
\beq
\begin{split}
    &\beta(\eta)=\frac{e^2+\frac{3}{2}g^2}{4\pi^2N\eta\Lambda^\epsilon}\cdot\eta,\\
	&\beta(e)=\left(\frac{\epsilon}{2}-\frac{e^2+\frac{3}{2}g^2}{4\pi^2 N\eta\Lambda^\epsilon}\right)\cdot e^2,\\
	&\beta(g^2)=\left(\frac{\epsilon}{2}-\frac{e^2+\frac{3}{2}g^2}{4\pi^2 N\eta\Lambda^\epsilon}\right)\cdot g^2.
\end{split}
\eeq
In terms of the dimensionless gauge couplings, $\tilde\alpha\equiv\frac{e^2}{4\pi^2\eta\Lambda^\epsilon},~
\alpha\equiv\frac{3g^2}{8\pi^2\eta\Lambda^\epsilon}$, the beta functions are
\beq
\begin{split}
&\beta(\tilde\alpha)=\left(\frac{\epsilon}{2}-\frac{\tilde\alpha+\alpha}{N}\right)\tilde\alpha,\\
&\beta(\alpha)=\left(\frac{\epsilon}{2}-\frac{\tilde\alpha+\alpha}{N}\right)\alpha.
\end{split}
\eeq

Before we end this appendix, we present some additional details for the calculation of the fermion self-energy, Eq. \eqref{eq: fermion propagator} (or Eq. \eqref{eq: fermion propagator U(2)}), given by
\beq
\begin{split}
	\Sigma_p(\omega, k_x, k_y)
	&=\int_{\Omega, q_x, q_y}(D_c(\Omega, q_y)+D_s(\Omega, q_y))G_{p}^{(0)}(\omega-\Omega, k_x-q_x, k_y-q_y)\\
	&=\int_{\Omega, q_x, q_y}(D_c(\Omega, q_y)+D_s(\Omega, q_y))\frac{1}{-i(\omega-\Omega)+p(k_x-q_x)+(k_y-q_y)^2}\\
	&=\int_{\Omega, q_y}\frac{i\cdot\sgn(\omega-\Omega)}{2}(D_c(\Omega, q_y)+D_s(\Omega, q_y)),
\end{split}
\eeq
where only the modes with $|q_y|<\Lambda$ should be integrated over.

Now we carry out the integral over $q_y$ for one of the two terms in the bracket \cite{mathbook}, and denote either $e_c^2$ or $e_s^2$ by $e^2$:
\beq
\int_{-\infty}^\infty\frac{dq_y}{2\pi}D(\Omega, q_y)=\frac{e^2}{\pi N}\int_0^\infty\frac{q_ydq_y}{\gamma e^2|\Omega|+q_y^{2+\epsilon}}=\frac{e^2}{\pi N}\cdot\frac{\pi}{2+\epsilon}\left(\gamma e^2|\Omega|\right)^{-\frac{\epsilon}{2+\epsilon}}\sin^{-1}\left(\frac{2\pi}{2+\epsilon}\right),
\eeq
and, for large $\Lambda\gg |\omega|$,
\beq
2\cdot\int_{\Lambda}^\infty\frac{dq_y}{2\pi}D(\Omega, q_y)=\frac{e^2}{\pi N}\int_\Lambda^\infty\frac{q_ydq_y}{\gamma e^2|\Omega|+q_y^{2+\epsilon}}\approx\frac{e^2}{\pi N}\int_{\Lambda}^\infty q_y^{-1-\epsilon}dq_y=\frac{e^2}{\pi N}\frac{\Lambda^{-\epsilon}}{\epsilon},
\eeq
so
\beq
\int_{-\Lambda}^\Lambda\frac{dq_y}{2\pi}D(\Omega, q_y)=\frac{e^2}{\pi N}\left[\frac{\pi}{2+\epsilon}\left(\gamma e^2|\Omega|\right)^{-\frac{\epsilon}{2+\epsilon}}\sin^{-1}\left(\frac{2\pi}{2+\epsilon}\right)-\frac{\Lambda^{-\epsilon}}{\epsilon}\right].
\eeq
Finally the integral over $\Omega$ leads to
\beq
\begin{split}
&\int\frac{d\Omega}{2\pi}\frac{i\cdot\sgn(\omega-\Omega)}{2}\cdot \frac{e^2}{\pi N}\left[\frac{\pi}{2+\epsilon}\left(\gamma e^2|\Omega|\right)^{-\frac{\epsilon}{2+\epsilon}}\sin^{-1}\left(\frac{2\pi}{2+\epsilon}\right)-\frac{\Lambda^{-\epsilon}}{\epsilon}\right]\\
=&\frac{ie^2}{4\pi N}(\gamma e^2)^{-\frac{\epsilon}{2+\epsilon}}\sin^{-1}\left(\frac{2\pi}{2+\epsilon}\right)|\omega|^{\frac{2}{2+\epsilon}}\sgn(\omega)-\frac{ie^2}{2\pi^2N}\cdot\frac{\Lambda^{-\epsilon}\omega}{\epsilon},
\end{split}
\eeq
which leads to the expressions for the above self-energy.

\section{BCS coupling} \label{app: BCS coupling}

For the sake of completeness, in this appendix we review the definition of the dimensionless BCS coupling \cite{Metlitski2014}. For this purpose, we will consider the physical case with $N=1$ and focus on the pairing between the two colors of the fermions. The dimensionless BCS coupling is extracted from the four-fermion interaction given by the following action: 
\beq
\begin{split}
S^{\BCS}
=&\int\prod_i^4\frac{d^2\vec k_id\omega_i}{(2\pi)^3}\psi^\dag_1(k_1)\psi^\dag_2(k_2)\psi_2(k_3)\psi_1(k_4)\cdot(2\pi)^3\delta^3\left(k_1+k_2-k_3-k_4\right)V(\vec k_1, \vec k_2; \vec k_3, \vec k_4)
\end{split}
\eeq
where $k=(\omega, \vec k)$ collectively denotes the frequency and momentum, and subscripts $1$ and $2$ are color indices. The patch indices are not shown explicitly and they can be inferred from the requirement of momentum conservation. Due to the kinematic constraints resulting from a Fermi surface, only couplings among the BCS-related momenta $V(\vec k_1, -\vec k_1; \vec k_2, -\vec k_2)$ are important \cite{SHANKAR1991, Polchinski1992, Shankar1993}. For simplicity, we assume that the system is rotationally invariant, such that the BCS couplings can be written as $V(\vec k_1, -\vec k_1; \vec k_2, -\vec k_2)=V(\theta_1-\theta_2)$, where the angles $\theta_{1,2}$ parameterize the contour of the Fermi surface. They can be further decomposed into angular harmonics:
\beq
V(\theta_1-\theta_2)=\sum_{m}V_me^{im(\theta_1-\theta_2)}.
\eeq
The dimensionless BCS coupling is then defined as
\beq \label{eq: dimensionless BCS coupling app}
\widetilde V_m=\frac{k_F}{2\pi v_F}V_m=\frac{k_F\eta}{2\pi}V_m,
\eeq
where $k_F$ is the Fermi momentum.

In the main text, we denote $\widetilde V\equiv \widetilde V_0$ to be the dimensionless BCS $s$-wave coupling constant.

\section{RG flows for the $U(1)\times U(1)$ gauge theory at $\epsilon=0$} \label{app: RG U(1) x U(1) epsilon=0}

Using the method in Ref. \cite{Metlitski2014}, in this appendix we discuss the flows described by Eq. \eqref{eq: beta functions epsilon=0 U(1) x U(1)}, which is reproduced here for clarity:
\beq \label{eq: RG U(1) x U(1) epsilon=0 app}
\begin{split}
    &\frac{d\alpha_c}{d\ell}=-\left(\alpha_c+\alpha_s\right)\cdot\alpha_c\\
    &\frac{d\alpha_s}{d\ell}=-\left(\alpha_c+\alpha_s\right)\cdot\alpha_s\\
    &\frac{d\widetilde V}{d\ell}=\alpha_c-\alpha_s-\widetilde V^2
\end{split}
\eeq
Notice that the physical value of $N=1$ has been substituted here.

Denoting $\alpha_t\equiv\alpha_c+\alpha_s$ and adding the first two equations in Eq. \eqref{eq: RG U(1) x U(1) epsilon=0 app} yield
\beq \label{eq: flow of sum of gauge couplings}
\frac{d\alpha_t}{d\ell}=-\alpha_t^2
\eeq
whose solution is given by
\beq
\alpha_t(\ell)=\frac{\alpha_t(0)}{1+\alpha_t(0)\ell}
\eeq
with $\alpha_t(0)$ a non-universal constant determined by microscopic details.

Recall from Appendix \ref{app: Wilsonian RG} that \beq
\frac{d}{d\ell}\left(\frac{\alpha_c}{\alpha_s}\right)=0
\eeq
holds exactly in this case, we set $r\equiv\frac{\alpha_c}{\alpha_s}$, which is also determined by non-universal microscopic details, and obtain the solutions for $\alpha_c$ and $\alpha_s$:
\beq
\begin{split}
    &\alpha_c(\ell)=\frac{\alpha_t(0)}{1+\alpha_t(0)\ell}\cdot\frac{r}{r+1}\\
    &\alpha_s(\ell)=\frac{\alpha_t(0)}{1+\alpha_t(0)\ell}\cdot\frac{1}{r+1}
\end{split}
\eeq

Next, we discuss the flow of $\widetilde V$. To solve for the flow of $\widetilde V$ when $r>1$, define $r_n\equiv\frac{r-1}{r+1}$, $\alpha_n\equiv \alpha_tr_n=\alpha_c-\alpha_s$ and $g\equiv\widetilde V/\sqrt{\alpha_n}$. Now solving for $\widetilde V$ amounts to solving for $g$,
and the flow equation of $g$ is
\beq
\frac{dg}{d\ell}=\sqrt{\alpha_n}(1-g^2)+\frac{g\alpha_n}{2r_n}
\eeq

As both $\alpha_n\rightarrow 0$, the last term of the above equation can be ignored. Then we get
\beq
\frac{dg}{d\alpha_n}=-\alpha_n^{-\frac{3}{2}}(1-g^2)r_n
\eeq
which yields
\beq
g(\ell)=\frac{\left(g(0)+1\right)e^{4r_n\left(\alpha_n^{-\frac{1}{2}}(\ell)-\alpha^{-\frac{1}{2}}_n(0)\right)}+g(0)-1}{\left(g(0)+1\right)e^{4r_n\left(\alpha_n^{-\frac{1}{2}}(\ell)-\alpha^{-\frac{1}{2}}_n(0)\right)}+1-g(0)}
\eeq

Physically, there are 3 distinct cases. 
\begin{enumerate}

\item When $g(0)>-1$, \ie $\widetilde V(0)>-\sqrt{\alpha_n(0)}$, $g\rightarrow 1$ as $\ell\rightarrow\infty$, that is,
\beq
\widetilde V(\ell)\rightarrow \sqrt{\alpha_n(\ell)}
\eeq
In this case, the FS is stable against pairing.

\item When $g(0)<-1$, \ie $\widetilde V(0)<-\sqrt{\alpha_n(0)}$, the FS is unstable, and $g(\ell)$, as well as $\widetilde V$, diverges at
\beq
\ell=r_n\left[\alpha_n^{-\frac{1}{2}}(0)+\frac{1}{4r_n}\ln\frac{\widetilde V(0)-\sqrt{\alpha_n(0)}}{\widetilde V(0)+\sqrt{\alpha_n(0)}}\right]^2-\alpha_t^{-1}(0)
\eeq

\item When $g(0)=-1$, \ie $\widetilde V(0)=-\sqrt{\alpha_n(0)}$, $g(\ell)=-1$, so
\beq
\widetilde V(\ell)=-\sqrt{\alpha_n(\ell)}=-\sqrt{\alpha_t(\ell)r_n}
\eeq
This is a transition line between the two other cases.

\end{enumerate}

Next, if $r<1$, $\widetilde V\rightarrow-\infty$ along the RG flow, and the FS is unstable. In this case, since $\alpha_n$ flows slowly, we treat it as a constant when analyzing the flow of $\widetilde V$, and obtain
\beq
\widetilde V(\ell)=\sqrt{-\alpha_n}\tan\left(-\sqrt{-\alpha_n}\ell+\tan^{-1}\frac{\widetilde V(0)}{\sqrt{-\alpha_n}}\right)
\eeq

Finally, if $r=1$, the contributions from the gauge fields to the RG equation for $\widetilde V$ in Eq. \eqref{eq: RG U(1) x U(1) epsilon=0 app} cancel, and the beta function of $\widetilde V$ to this order is the same as that of a Fermi liquid:
\beq \label{eq: Fermi liquid form}
\frac{d\widetilde V}{dl}=-\widetilde V^2.
\eeq
Therefore, to analyze the effect of gauge fields, we need to go to the next-leading order, and we will see that the result at this next-leading order further suppress pairing. To this end, it is more convenient to to consider the effective Lagrangian given by Eqs. \eqref{eq: 2-U(1) patch}, \eqref{eq: 2-U(1) patch fermion} and \eqref{eq: gauge fields action U(1) x U(1)}, but in terms of $a_{1,2}$, rather than $a_{c,s}$:
\beq
\mc{L} = \mc{L}_f + \mc{L}_{[a_1, a_2]},
\eeq
with
\beq
\begin{split}
\mc{L}_f=
\sum_{p=\pm,\alpha=1,2}\psi^\dagger_{\alpha p}\left[\eta\partial_\tau + v_F\left(-ip\partial_x -\partial_y^2\right)\right]\psi_{\alpha p}- v_F \sum_{p=\pm}\lambda_p\left(a_1\psi_{1p}^\dag \psi_{1p}+a_2\psi_{2p}^\dag \psi_{2p}\right),
\end{split}
\eeq
and
\beq \label{eq: patch Lagrangian in terms a1 and a2}
\mc{L}_{[a_1, a_2]}=\frac{N}{4e^2}|k_y|^{1+\epsilon}\left(|a_1|^2+|a_2|^2\right),
\eeq
with $e=e_c=e_s$. Importantly, when $r=1$ and $\epsilon<1$, there is no direct coupling between $a_1$ and $a_2$ in this effective Lagrangian, and such a coupling cannot be generated in the RG process due to the non-analytic nature of Eq. \eqref{eq: patch Lagrangian in terms a1 and a2}. This is actually another way to see why the contribution from the gauge fields to the flow of $\widetilde V$ vanishes at the leading order, given by Eq. \eqref{eq: Fermi liquid form}.

\begin{figure}[h]
\begin{center}
\includegraphics[scale=0.2]{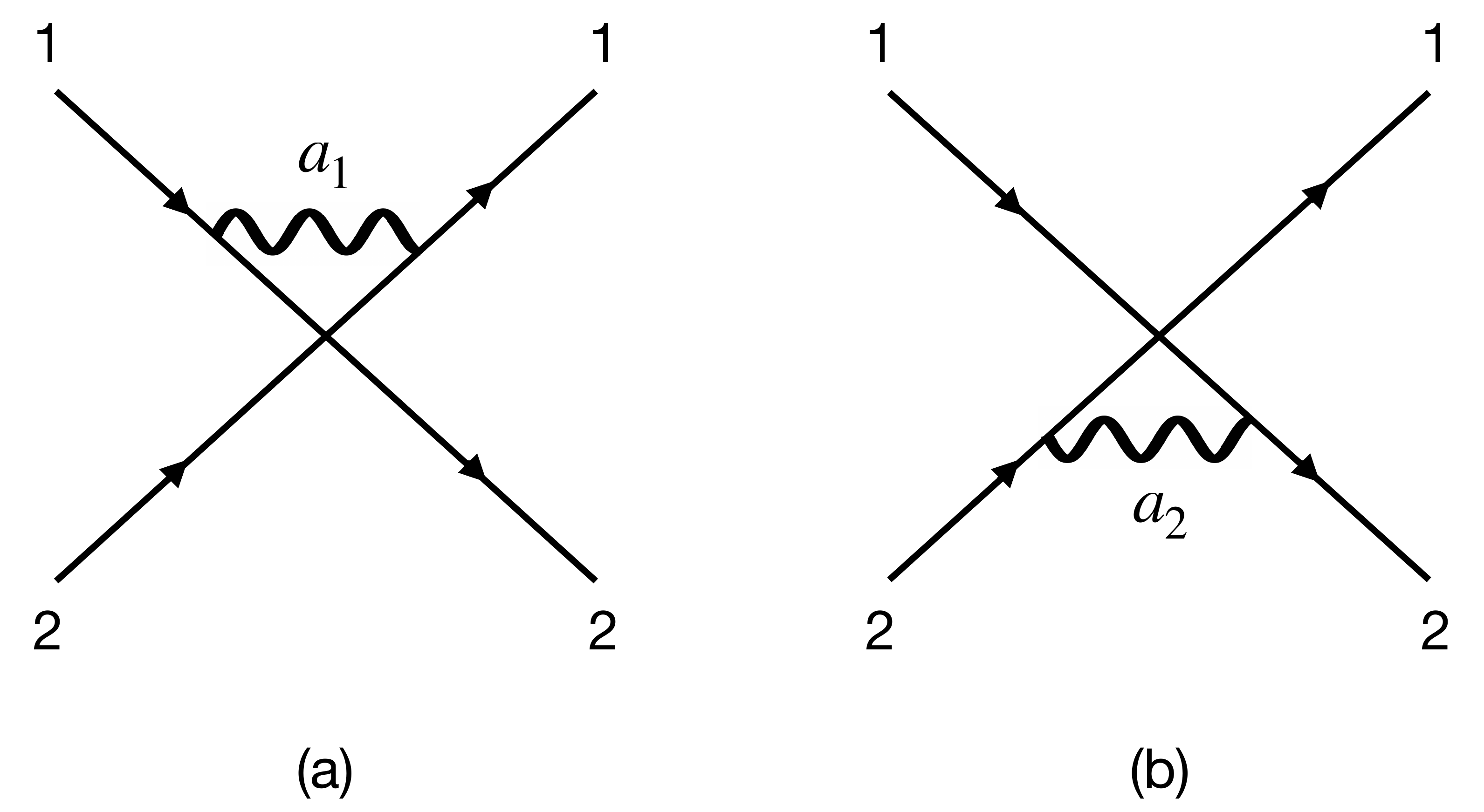}
\end{center}
\caption{The next-leading-order contribution to the flow of $\widetilde V$ coming from vertex correction. Notice when $r=1$, there is no gauge field line that connects the fermions with the two different colors.}
\label{pairvertex}
\end{figure}

The next-leading order contribution comes from vertex correction in Fig.~\ref{pairvertex} and from the flow of $\eta$ in the definition of 
$\widetilde V$, Eq. \eqref{eq: dimensionless BCS coupling app} \cite{Metlitski2014}. This vertex correction vanishes as a result of the pole-structure of the fermionic propagators \cite{Mross2010}. Notice that there is no analogous vertex correction with an internal gauge field line connecting two fermion lines with two different colors (i.e. connecting `1' and `2'), because of the absence of a coupling between $a_1$ and $a_2$ when $r=1$. The physical implication of the flow of $\eta$ is that quasiparticles are destroyed, which is expected to further suppress pairing. Indeed, according to Eqs. \eqref{eq: beta functions skip point} and \eqref{eq: dimensionless BCS coupling app}, the flow of $\eta$ changes the beta function of $\widetilde V$ into
\beq
\frac{d\widetilde V}{dl}=\alpha_t\widetilde V-\widetilde V^2.
\eeq
The first term, coming from the flow of $\eta$, further suppresses pairing. In the above equation, because $\alpha_t$ flows slowly, we view it as a constant, and obtain a ``fixed point" of $\widetilde V$ at $\widetilde V_*=\alpha_t$ if $\widetilde V(0)>0$, which then flows to zero according to Eq. \eqref{eq: flow of sum of gauge couplings} {\footnote{There can be higher order corrections that can modify how exactly $\alpha_t$ flows to zero, but they are not expected to change the result that $\alpha_t$ will slowly flow to zero.}}. In this case, the FS is stable against pairing. If $\widetilde V(0)<0$, $\widetilde V\rightarrow-\infty$ along the RG flow, which means the FS is unstable to pairing. Taken together, we conclude that when $r=1$, gauge fields suppress pairing, and the FS are perturbatively stable.

In summary, when $r<1$, the FS is unstable to pairing, while it is stable when $r\geqslant 1$.

\end{document}